# Stochastic assignment games for Mobility-as-a-Service markets

Bingqing Liu[1*], David Watling[2], Joseph Y. J. Chow[3]

[1]Mobility Center of Excellence on New Mobility and Automated Vehicles, University of California, Los Angeles, Los Angeles, CA 90095, US

[2]Institute for Transport Studies, University of Leeds, Woodhouse, Leeds LS2 9JT, UK

[3]C2SMARTER University Transportation Center, NYU Tandon School of Engineering, New York, NY 10012, US

[*]Corresponding author email: bingqingliu777@g.ucla.edu

**Abstract**

We study the stochastic assignment game and extend it to model multimodal mobility markets with a regulator or a Mobility-as-a-Service (MaaS) platform. We start by presenting general forms of one-to-one and many-to-many stochastic assignment games. Optimality conditions are discussed. The core of stochastic assignment games is defined, with *expected payoffs* of sellers and buyers in stochastic assignment games as payoffs from a hypothetical "ideal matching" that represent sellers' and buyers' expectations under imperfect information. To apply stochastic assignment games to the urban mobility markets, we extend the general stochastic many-to-many assignment game into a stochastic Stackelberg game to model MaaS systems, where the platform is the leader, and users and operators are the followers. The platform sets fares to maximize revenue. Users and operator react to the fare settings to form a stochastic many-to-many assignment game considering both fixed-route services and Mobility-on-Demand (MOD). The Stackelberg game is formulated as a bilevel problem. The lower level is the stochastic many-to-many assignment game between users and operators, shown to yield a coalitional logit model. The upper-level problem is a fare adjustment problem maximizing revenue. An iterative balancing algorithm is proposed to solve the lower-level problem exactly. The bilevel problem is solved through an iterative fare adjusting heuristic, whose solution is shown to be equivalent to the bilevel problem with an additional condition when it converges. Two case studies are conducted. The model can be applied to design MaaS fares maximizing income of the platform while anticipating the selfish behavior and heterogeneity of users and operators. Public agencies can also use the model to manage multimodal transportation systems.

## 1 Introduction

Development of information and communication technologies (ICT) is giving rise to new mobility services, such as bikesharing, micromobility, carsharing, ride-hailing, microtransit, and peer-to-peer ridesharing. The coexistence of mobility services with higher flexibility and traditional mass transit leads to the concept of Mobiltiy-as-a-Service (MaaS), which provides mobility services through a joint digital channel that enables users to plan, book, and pay for multiple types of mobility services (Wong et al., 2020; van den Berg et al., 2022). We refer to these joint digital channels as "platforms" not simply as software, but in the sense of platform economies (Kenney and Zysman, 2016). On the one hand, with different types of mobility services incorporated, MaaS systems can meet more diverse, multimodal travel demand with higher spatial and temporal flexibility, providing a better travel experience and doing so with higher efficiency compared with mobility markets without an integrator like the MaaS platform. On the other hand, such systems are more complicated to model, manage, and evaluate.

Evaluation of MaaS platforms does not depend only on the behaviors of travelers, i.e. how they respond to additional route options that may include transfers. Rather, it needs to also incorporate the supply-side behavior of the operators in their interactions with travelers and each other, as well as the role of the platform. Mobility markets with multiple operators are mostly modeled with noncooperative games and simulations. Derks and Tijs (1985) modelled such markets with noncooperative games between operators without travelers' route choices. Zhou et al. (2005) modelled operators with noncooperative games and users with a stochastic user equilibrium (SUE), forming a Stackelberg game with the operators as the leader and travelers as the followers. Zhou et al. (2022) represented competitive ride-sourcing operators with single-operator trips (unimodal trips). Najmi et al. (2023) also modelled operators with noncooperative games but incorporating multimodal trips.

An alternative approach to modeling MaaS systems is through an assignment game, considering users as the "buyers" and operators as "sellers", forming a two-sided market within this platform. An assignment game (Shapley and Shubik, 1971) is a special case of a stable matching problem (Gale and Shapley, 1962), a type of cooperative game in which utilities are transferable between buyers and sellers and a match is determined such that no one has incentive to break from their match. The game is defined by a matching problem and its corresponding stability conditions. The assignment game has been adapted to model transportation services as a many-to-many assignment game. Rasulkhani and Chow (2019) modelled MaaS systems as a many-to-one assignment game such that multiple users were matched to unimodal route. Pantelidis et al. (2020) proposed a many-to-many assignment game model such that multiple users were matched to multimodal paths on which links are owned by different operators. In that model, the matching problem was a multicommodity flow network design problem where users decide paths to take while operators decide which links to operate.

The many-to-many assignment game has been shown by Pantelidis et al. (2020) to model both cooperative and competitive behavior between operators, where the cost allocation bound represents the threshold separating cooperative sharing of profits from competitive outcomes. However, that model does not recognize Mobility-on-Demand (MOD) operators and their unique characteristics. Such on-demand services cover all OD pairs within a service region and steady state wait times depend on both the demand and fleet size. Liu and Chow (2024) further expanded the many-to-many assignment model to incorporate such MOD characteristics using macroscopic meeting cost functions and proposing a stability guarantee through subsidy provision. Xi et al. (2022) also modelled MaaS as two-sided markets with fixed-route transit and MOD, considering single-operator trips assuming a specific cost allocation mechanism. Yao and Bekhor (2023) modelled ridesharing markets with multiple operators with stable matching, not incorporating fixed-route transit services.

In additional to users and operators, the platform is also a decision-maker of MaaS systems, who regulates the market in which the matchings are formed. The role of the operator is not explored comprehensively in the literature. Xi et al. (2023) proposed an auction-based mechanism and tractable optimization models for the MaaS platform to manage the demand, not considering operator behavior. Ding et al. (2023) introduced offline and online mechanisms of matching and pricing which endogenously infers users' trip utilities through an auction-based method, also ignoring operators' choices. Liu and Chow (2024) modelled the role of the platform as subsidy provider to stabilize unstable matchings while considering operators' behavior. Yao and Zhang (2024) adopted a similar framework as Liu and Chow (2024). The key difference is that the latter uses fixed-path pricing and binary variables for service capacity, while the former introduces OD-based pricing, and explicitly models the interactions between MaaS and non-MaaS systems with shared congestion effects and equilibrium constraints within the multi-modal traffic network.

In the many-to-many assignment framework (Pantelidis et al., 2020, Liu and Chow, 2024; Yao and Zhang, 2024), the matching is not dependent on the fare setting. Fares that stabilize the matching is not determined by fare setting. Rather, the model first solves the assignment problem, and then derives stabilizing fares from the cost allocation, which distributes surplus consistently across users and operators. This can be interpreted as a special case Stackelberg game where the upper-level problem of fare design from the platform's perspective (leader) does not affect the lower-level matching problem between operators and users (followers). The intuition is like how first-best marginal cost pricing can be set by first determining the system optimal assignment before setting the optimal price to incentivize travelers toward that state (Yang and Huang, 1998). However, in the application we consider, such an assumption heavily restricts the range of fare setting, sometimes leading to infeasibility of the upper-level problem without further consideration of platform subsidies (Liu and Chow, 2024).

Furthermore, an important research gap is related to the heterogeneity that exists among both users and operators, which has not been directly captured in the current MaaS literature. In practice, users differ in their sensitivity to travel time, monetary cost, and waiting time, leading them to form different perceptions of the same travel conditions. Operators also face heterogeneous cost structures due to differences in situations such as vehicle procurement, depreciation methods, maintenance procedures, driver wage arrangements, and dispatch strategies. These factors influence how users and operators evaluate alternatives and react to fares. Such heterogeneity also raises further conceptual questions: Who are the different types of decision-makers making choices when heterogeneity and interactions are present? How should the stability conditions be modeled when the matching outcome follows a stochastic assignment framework rather than a deterministic one? Incorporating these considerations requires a stochastic representation in which decisions are made probabilistically under imperfect information. This leads to a stochastic matching equilibrium rather than an optimization of a system-wide objective.

We aim to address the research gaps by proposing a stochastic assignment game framework. We start by presenting general forms of one-to-one and many-to-many stochastic assignment games, in which the decision variables are the probabilities that matchings are formed. Optimality conditions are discussed. The core of a stochastic assignment game is defined in which *expected payoffs* of sellers and buyers in stochastic assignment games are from a hypothetical "ideal matching" that represent sellers' and buyers' expectations under imperfect information. To apply stochastic assignment games to the urban mobility markets, we extend the general stochastic many-to-many assignment game into a stochastic multimodal assignment game. Based on that, we propose a stochastic multimodal Stackelberg game in which the platform is the leader, and the coalitions formed between users and operators are the followers, considering both fixed-route and MOD operators. In this way, the decision-makers choosing routes are the potential coalitions of operators and users; price setting is an upper-level platform decision that directly impacts those choices. Users and operators react to the fares, forming many-to-many matchings. The many-to-many assignment game considers the stochasticity on both users' and operators' sides and incorporates the impact of fares on the matchings. The lower-level problem is shown to yield a coalitional logit model of users and operators, as a "stochastic matching equilibrium" model, and an iterative balancing algorithm (Bell, 1995) is proposed to solve it. The bilevel problem is solved through an iterative fare adjusting algorithm, whose solution is shown to be equivalent to the original Stackelberg game problem when it converges. Two case studies are conducted to serve complementary purposes: the small illustrative example clarifies the mechanics of the proposed framework, while the larger case study demonstrates its scalability and real-world applicability.

## 2 Stochastic Assignment Game

As discussed by Liu and Chow (2024), deterministic assignment games solve for deterministic matchings through binary variables (i.e., 0–1 decisions on whether a match occurs), assuming players have perfect information about the value of each possible match. However, players often face uncertainty about costs, benefits, or even the willingness of counterparts to form a match, so the perfect-information assumption rarely holds. Stochastic assignment games introduce uncertainty into the payoffs and solve probabilities of different matchings, which allows the model to capture market volatility and incomplete information.

We start from general formulations of the one-to-one stochastic assignment game, many-to-one stochastic game, and many-to-many stochastic assignment game, developing the optimality conditions. Then, we extend the stochastic many-to-many assignment game to a stochastic multimodal assignment game to model urban mobility markets from a regulatory perspective. A stochastic multimodal Stackelberg game is formulated to solve for optimal fares and flows in which the regulator/platform is the leader, and travelers and operators are the followers. Solution algorithms are proposed, and their convergence properties are analyzed.

### 2.1 One-to-one stochastic assignment game

We adopt the notation from Shapley and Shubik (1971) to formulate the one-to-one stochastic assignment game denoted by $P_1$. There are 2 sets of players, sellers indexed by $i \in Q$ and buyers indexed by $j \in P$. Seller $i$ values his/her product at $c_i$. Buyer $j$ values the product of seller $i$ at $h_{ij}$. These are numerical representations of buyer and seller valuations in a common unit (e.g. dollars). When buyer $j$ and seller $i$ form a matching, buyer $j$ pays a price $p_i$ to seller $i$, and the product of seller $i$ is transferred from seller $i$ to buyer $j$. In this process, the gain of seller $i$ is $p_i - c_i$, and the gain of buyer $j$ is $h_{ij} - p_i$. An example of such a one-to-one assignment game includes the private home market (Shapley and Shubik, 1971).

The *characteristic function* of the game, which states the worth $a_{ij}$ of a matching $(i,j)$ is the sum of the gain of seller $i$ and buyer $j$. In deterministic assignment games, the characteristic function is $a_{ij} = \max(0, h_{ij} - c_i)$, where as above $h_{ij}$ is buyer $j$'s valuation of seller $i$'s product, and $c_i$ is seller $i$'s valuation of his/her product. This computation assumes that a matching will not form if there is no positive gain to be shared by the seller and buyer. However, in the stochastic case with imperfect information, that is no longer the case. Any matching should have a non-negative probability of forming. Hence in stochastic assignment games, we write the characteristic function as $a_{ij} = h_{ij} - c_i$.

$$P_1: \min_{x} \sum_{i \in Q} \sum_{j \in P} x_{ij}(\ln(x_{ij}) - 1) - \alpha \sum_{i \in Q} \sum_{j \in P} a_{ij} x_{ij} \tag{1a}$$

Subject to

$$\sum_{j \in P} x_{ij} \leq 1, \quad \forall i \in Q \quad (v_i) \tag{1b}$$

$$\sum_{i \in Q} x_{ij} \leq 1, \quad \forall j \in P \quad (u_j) \tag{1c}$$

$$x_{ij} \geq 0, \quad \forall i \in Q, j \in P \tag{1d}$$

In stochastic assignment games, the decision variables $x_{ij}$ are continuous variables between 0 and 1 indicating the probability that a matching $(i,j)$ is formed, which differs from the binary variables in the deterministic assignment games. In the stochastic matching problem $P_1$, the objective function Eq.(1a) is the sum of the total payoff and an entropy term $\sum_{i \in Q} \sum_{j \in P} x_{ij}(\ln(x_{ij}) - 1)$. The entropy term leads to stochastic matchings that conform to a logit model as discussed in **Theorem 1**. The parameter $\alpha > 0$ reflects the level of stochasticity: the level of stochasticity decreases as $\alpha$ increases. Due to the entropy term, $x_{ij}$ is strictly positive. Constraints Eqs. (1b) – (1c) ensure that the total probability of forming matchings for each player $i$ or $j$ is less than or equal to 1. The Lagrange multipliers of constraints Eq. (1b) and Eq. (1c) are $v_i$ and $u_j$.

The core of stochastic assignment games is introduced in **Theorem 1, which defines** *expected payoffs* of sellers and buyers in stochastic assignment games. To do so, we first define an *ideal matching*: a deterministic assignment that would arise under complete information, in which each seller–buyer pair forms whenever their joint valuation exceeds the seller's cost. This benchmark outcome occurs with probability 1, and serves as the reference against which expectations under imperfect information are computed. Building on this notion, Theorem 1 establishes that the expected payoffs in stochastic assignment games correspond to those derived from this ideal matching.

**Theorem 1.** Core of stochastic assignment game. *The core of a stochastic assignment game is the set $\{v_i/\alpha, \forall i \in Q; u_j/\alpha, j \in P\}$, where $v_i/\alpha$ and $u_j/\alpha$ are the **expected payoffs** of seller $i \in Q$ and buyer $j \in P$ which are payoffs from an hypothetical "ideal matching" with probability 1. Sum of $v_i/\alpha$ and $u_j/\alpha$ is always larger than the total payoffs generated from the matching $(i,j)$ which is equal to $a_{ij}$.*

***Proof***. Since $P_1$ is non-linear, we write the Lagrangian of $P_1$. The Lagrangian is written as Eq. (2). Taking the first derivative of $L$ w.r.t. $x_{ij}$ leads to Eq. (3).

$$L = \sum_{i \in Q} \sum_{j \in P} x_{ij}(\ln(x_{ij}) - 1) - \alpha \sum_{i \in Q} \sum_{j \in P} a_{ij} x_{ij} + \sum_{i \in Q} v_i (\sum_{j \in P} x_{ij} - 1)$$
$$+ \sum_{j \in P} u_j (\sum_{i \in Q} x_{ij} - 1) \quad (2)$$

$$\frac{\partial L}{\partial x_{ij}} = \ln(x_{ij}) - \alpha a_{ij} + v_i + u_j \quad (3)$$

The KKT conditions of $P_1$ are Eqs. (4) – (8) and Eqs. (1b) – (1c).

$$\ln(x_{ij}) - \alpha a_{ij} + v_i + u_j = 0, \quad \forall i \in Q, j \in P \quad (4)$$

$$v_i \left( \sum_{j \in P} x_{ij} - 1 \right) = 0, \quad \forall i \in Q \quad (5)$$

$$u_i \left( \sum_{i \in Q} x_{ij} - 1 \right) = 0, \quad \forall j \in P \quad (6)$$

$$v_i \geq 0, \quad \forall i \in Q \quad (7)$$
$$u_i \geq 0, \quad \forall j \in P \quad (8)$$

Eq. (4) reduces to Eq. (9).

$$\ln(x_{ij}) = \alpha a_{ij} - v_i - u_j, \quad \forall i \in Q, j \in P \quad (9)$$

In a deterministic assignment game, if $a_{ij} = {v_i}/{\alpha} + {u_j}/{\alpha}$, $x_{ij}^* = 1$, then the probability of matching $(i,j)$ is 1. ${v_i}/{\alpha}$ and ${u_j}/{\alpha}$ are the payoffs of an "ideal matching" $(i,j)$. However, in logit stochastic assignment games with imperfect information, we always have $0 < x_{ij}^* < 1$, since $x_{ij}^*$ is strictly positive for any matching $(i,j)$. This means that $a_{ij} < {v_i}/{\alpha} + {u_j}/{\alpha}$ always holds. Intuitively, $\alpha$ acts as a scale parameter controlling the relative importance of deterministic valuations versus random perturbations. Dividing by $\alpha$ rescales the payoffs into a common utility scale that reflects the degree of uncertainty. We name ${v_i}/{\alpha}$ and ${u_j}/{\alpha}$ as the *expected payoffs* of $i$ and $j$. The actual total payoffs generated from a matching $(i,j)$, which is $a_{ij}$, is always smaller than the sum of expected payoffs.

∎

We define the right-hand side of Eq. (9) as the utility of matching $(i,j)$, denoted by $u_{ij}$. We refer to it as the **utility**. Holding other factors constant, higher $u_{ij}$ leads to higher probability $x_{ij}$. The following choice probabilities in Eqs. (10) – (11) can be derived from Eq. (9).

$$x_{ij} = \frac{\exp(u_{ij})}{\sum_{j' \in P} \exp(u_{ij'})} = \frac{\exp(\alpha a_{ij} - u_j)}{\sum_{j' \in P} \exp(\alpha a_{ij'} - u_{j'})} \quad (10)$$

$$x_{ij} = \frac{\exp(u_{ij})}{\sum_{i' \in Q} \exp(u_{i'j})} = \frac{\exp(\alpha a_{ij} - v_i)}{\sum_{i' \in Q} \exp(\alpha a_{i'j} - v_{i'})} \quad (11)$$

We use the following numerical example, taken from Shapley and Shubik (1971) with valuations changed, to illustrate the one-to-one stochastic assignment game. Let there be three sellers (1,2,3) and three buyers (1',2',3') as shown in **Table 1**.

Table 1. Sellers' and buyers' valuations of the products.

| Product ($i$) | Sellers' valuations ($c_i$) | Buyers' valuations | | |
|---|---|---|---|---|
| | | ($h_{i1}$) | ($h_{i2}$) | ($h_{i3}$) |
| 1 | 37 | 42 | 41 | 42 |
| 2 | 25 | 26 | 23 | 25 |
| 3 | 43 | 47 | 48 | 46 |

The characteristic functions $a_{ij}$ can be computed accordingly, shown in **Table 2**. The first column indicates seller index $i$ and the top row indicates buyer index $j$.

Table 2. Characteristic functions derived from valuations.

| $a_{ij}$ | 1' | 2' | 3' |
|---|---|---|---|
| 1 | 5 | 4 | 5 |
| 2 | 1 | -2 | 0 |
| 3 | 4 | 5 | 3 |

Assuming $\alpha = 1$, the solution is shown in **Table 3**.

Table 3. Probabilities of matchings and expected payoffs.

| $x_{ij}^*$ | 1' | 2' | 3' | $v_i/\alpha$ |
|---|---|---|---|---|
| 1 | 0.285 | 0.195 | **0.520** | 3.763 |
| 2 | **0.567** | 0.053 | 0.381 | -0.925 |
| 3 | 0.148 | **0.752** | 0.100 | 3.415 |
| $u_j/\alpha$ | 2.492 | 1.870 | 1.891 | |

Eqs. (10) – (11) are verified in **Table 4**.

Table 4. Choice model verification.

| Matching ($i,j$) | Probability $x_{ij}^*$ | Utility $u_{ij}$ | $\frac{\exp(u_{ij})}{\sum_{j' \in P} \exp(u_{ij'})}$ | $\frac{\exp(u_{ij})}{\sum_{i' \in Q} \exp(u_{i'j})}$ |
|---|---|---|---|---|
| (1,1') | 0.296 | -1.256 | 0.296 | 0.296 |
| (1,2') | 0.195 | -1.633 | 0.195 | 0.195 |
| (1,3') | 0.520 | -0.654 | 0.520 | 0.520 |
| (2,1') | 0.567 | -0.568 | 0.567 | 0.567 |
| (2,2') | 0.053 | -2.945 | 0.053 | 0.053 |
| (2,3') | 0.381 | -0.966 | 0.381 | 0.381 |
| (3,1') | 0.148 | -1.908 | 0.148 | 0.148 |
| (3,2') | 0.752 | -0.285 | 0.752 | 0.752 |
| (3,3') | 0.100 | -2.306 | 0.100 | 0.100 |

If we solve the example as a deterministic assignment game, the solution is: $x^*_{13'} = 1$, $x^*_{21'} = 1$, $x^*_{32'} = 1$, with everything else as 0. The non-zero $x^*_{ij}$ correspond to the largest probabilities in each column and each row of **Table 3**.

## 2.2 Many-to-one stochastic assignment game

Based on the stochastic one-to-one assignment game, we can formulate the a stochastic many-to-one assignment game. In a many-to-one assignment game, the set of sellers $Q$ can be matched with one or more buyers with a limit of matching seller $i$ with at most $w_i$ buyers (for any $i \in Q$). Rasulkhani and Chow (2019) proposed a many-to-one assignment game to model a unimodal traffic equilibrium with joint traveler and operator behavior, which is the deterministic case of the many-to-one assignment game. In that case, buyers $j \in P$ are travelers, and sellers $i \in Q$ are routes. Each traveler can choose only one route while a route can be chosen by more than one traveler. Parameter $w_i$ is the capacity of route $i$, and Eq. (12b) is the capacity constraint. Dual variable $v_i$ of the capacity constraint can be interpreted as the delay caused by limited capacity. At the same time, $u_i$ is interpreted as the payoff of seller $i$ (route $i$), indicating that the payoff of the seller $i$ in a matching $(i, j)$ can be interpreted as the loss of buyer $j$.

In the stochastic case, an entropy term is added to the objective (Eq. (12a)), and $x_{ij}$ is the non-zero probability of the matching $(i, j)$. The left-hand-side of constraint Eq. (12b) is the *expected* number of buyers matched with seller $i$, which is smaller than or equal to its capacity.

$$P_2: \min_x \sum_{i \in Q} \sum_{j \in P} x_{ij}(\ln(x_{ij}) - 1) - \alpha \sum_{i \in Q} \sum_{j \in P} a_{ij} x_{ij} \tag{12a}$$

Subject to

$$\sum_{j \in P} x_{ij} \leq w_i, \quad \forall i \in Q \quad (v_i) \tag{12b}$$

$$\sum_{i \in Q} x_{ij} \leq 1, \quad \forall j \in P \quad (u_j) \tag{12c}$$

$$x_{ij} \geq 0, \quad \forall i \in Q, j \in P \tag{12d}$$

We write the Lagrangian (Eq. (13)) and take the derivative (Eq. (14)). The solution of the optimality condition is shown in Eq. (15).

$$L = \sum_{i \in Q} \sum_{j \in P} x_{ij}(\ln(x_{ij}) - 1) - \alpha \sum_{i \in Q} \sum_{j \in P} a_{ij} x_{ij} + \sum_{i \in Q} v_i (\sum_{j \in P} x_{ij} - w_i)$$
$$+ \sum_{j \in P} u_j (\sum_{i \in Q} x_{ij} - 1) \tag{13}$$

$$\frac{\partial L}{\partial x_{ij}} = \ln(x_{ij}) - \alpha a_{ij} + v_i + u_j \tag{14}$$

$$\ln(x_{ij}) = \alpha a_{ij} - v_i - u_j \tag{15}$$

Eq. (15) is the same as Eq. (9). **Theorem 1** is applicable to the many-to-one assignment game. The only difference is that $u_i$ is the expected payoff that seller $i$ gains from one buyer $j$ from the matching $(i, j)$, and the total expected payoff of seller $i$ becomes $w_i u_i$.

In traffic assignment, travelers from the same OD pair are modeled as homogenous groups, each with a route set to choose from. In such cases, sellers are the routes and buyers are the traveler groups. We use $S$ to denote the traveler groups and $d_s, s \in S$ to denote the number of travelers of each group. Decision variables are $f_{rs}$, which is the flow on route $r$ from group $s$. We assume that travelers of group $s \in S$ gain utility $U_s$ upon completing their trip. The worth of a traveler in group $s$ making the trip using path $r$ is $a_{rs} = U_s - t_{rs}$ (excluding operation cost), where $t_{rs}$ is the travel cost of a traveler in group $s$ on path $r$. $P_2$ is modified into $P_3$ (Eq. (16)), which is equivalent to the capacitated stochastic user equilibrium proposed by Bell (1995).

$$P_3: \min_f \sum_{s \in S} \sum_{r \in R} f_{rs}(\ln(f_{rs}) - 1) + \alpha \sum_{s \in S} \sum_{r \in R} t_{rs} f_{rs} \tag{16a}$$

Subject to

$$\sum_{s \in S} f_{rs} \leq w_r, \quad \forall r \in R \quad (v_r) \tag{16b}$$

$$\sum_{r \in R} f_{rs} \leq d_s, \quad \forall s \in S \quad (u_s) \tag{16c}$$

$$f_{rs} \geq 0, \quad \forall r \in R, s \in S \tag{16d}$$

Bell (1995) showed that the optimality condition of $P_3$ is Eq. (17) and proved that $v_r/\alpha$ is equivalent to the delay when the capacity of route $r$ is binding. The term $v_r/\alpha$ can be interpreted as both the traveler's delay on route $r$ and route $r$'s expected payoff per traveler. In a matching, one party's payoff can be interpreted as the other party's loss.

$$\ln(f_{rs}) = -\alpha t_{rs} - v_r - u_s, \quad \forall r \in R, s \in S \tag{17}$$

## 2.3 Many-to-many stochastic assignment game

Pantelidis et al. (2020) extended the many-to-one assignment game proposed by Rasulkhani and Chow (2019) into a many-to-many assignment game to model multimodal traffic assignment with joint traveler and operator behavior. Here we define a general form of the many-to-many assignment game. Sellers $i \in Q$ and buyers $j \in P$ are both allowed to be matched with multiple counterparts. We define a multi-matching $m$ of buyer $j$ as a combination of multiple single matchings between a buyer $j \in P$ and sellers in $Q$, which can be denoted as $m \in M_j$, where $M_j = \{(i,j), \forall i \in Q_{mj}\}$ and $Q_{mj} \subseteq Q$ is the set of sellers matched with buyer $j$ in $m$. Unlike the many-to-many assignment defined by Sotomayor (1992), we study the type of many-to-many assignment game in which each multi-matching $m$ of buyer $j$ generates a value $a_{mj}$, which is allocated between buyer $j$ and all relevant sellers $i \in Q_{mj}$. We assume that all possible multi-coalitions of buyers $j \in P$ can be enumerated to form a set $M_j$, and the coalitions formed by buyer $j$ is exactly one multi-coalition from the enumeration. Seller $i \in Q$ has a limit of matching with at most $w_i$ buyers, which is reflected as constraint Eq. (18b). Let the parameter $\delta_{imj}$ be a binary indicator that equals 1 when $i \in Q_{mj}$, 0 otherwise. Decision variable $x_{mj}$ is the probability that multi-matching $m \in M_j$ is formed by buyer $j \in P$. We have Lagrange multipliers $v_i$ and $u_j$ corresponding to constraints Eq. (18b) and Eq. (18c).

$$P_4: \min_x \sum_{j \in P} \sum_{m \in M_j} x_{mj}(\ln(x_{mj}) - 1) - \alpha \sum_{j \in P} \sum_{m \in M_j} a_{mj} x_{mj} \tag{18a}$$

Subject to

$$\sum_{j \in P} \sum_{m \in M_j} \delta_{imj} x_{mj} \leq w_i, \quad \forall i \in Q \quad (v_i) \tag{18b}$$

$$\sum_{m \in M_j} x_{mj} \leq 1, \quad \forall j \in P \quad (u_j) \tag{18c}$$

$$x_{mj} \geq 0, \quad \forall j \in P, m \in M_j \tag{18d}$$

Similarly, we write the Lagrangian as Eq. (19) and take its derivative Eq. (20).

$$\begin{aligned} L &= \sum_{j \in P} \sum_{m \in M_j} x_{mj}(\ln(x_{mj}) - 1) - \alpha \sum_{j \in P} \sum_{m \in M_j} a_{mj} x_{mj} \\ &\quad + \sum_{i \in Q} v_i \Bigl( \sum_{j \in P} \sum_{m \in M_j} \delta_{imj} x_{mj} - w_i \Bigr) + \sum_{j \in P} v_j \Bigl( \sum_{m \in M_j} x_{mj} - 1 \Bigr) \end{aligned} \tag{19}$$

$$\frac{\partial L}{\partial x_{mj}} = \ln(x_{mj}) - \alpha a_{mj} + \sum_{i \in Q} \delta_{imj} v_i + u_j \tag{20}$$

Eq. (20) reduces to Eq. (21) with the KKT conditions.

$$\ln(x_{mj}) = \alpha a_{mj} - \sum_{i \in Q} \delta_{imj} v_i - u_j, \quad \forall i \in Q, j \in P \tag{21}$$

**Theorem 1** is still applicable to stochastic many-to-many assignment games. In this case, $v_i/\alpha$ is the expected payoff of seller $i$, and $\sum_{i \in Q} \delta_{imj} v_i/\alpha$ is the total expected payoff of all the sellers involved in the multi-matching $m \in M_j$ of buyer $j \in P$.

Consider a multimodal network defined as a directed graph $G(N, A)$ serving a set of traveler groups $S$ traveling to/from centroids $N_Z \subset N$. Travelers in one traveler group $s \in S$ are composed of a homogeneous population of travelers $d_s$ with the same origin-destination (OD) pair. Each link $l \in A$ is associated with a capacity $w_l$, an operating cost per traveler $\bar{c}_l$, and a travel cost $t_l$. Similar to $P_3$, we use continuous variables $f_{rs}$ to denote the number of travelers in group $s$ that choose path $r$. The path set of group $s \in S$ is assumed to be enumerated as set $R_s$. $\delta_{lrs} = 1$ indicates that link $l \in A$ is on path $r \in R_s$ of group $s \in S$. We assume that travelers of traveler group $s \in S$ gain utility $U_s$ upon completing their trip. The value of $U_s$ is typically dependent on trip purpose. Ma et al. (2021) provided an example of calibrating these utilities from existing travel modes. Assuming $U_s$ is large enough, the worth of a traveler in group $s$ making the trip using path $r$ is $a_{rs} = U_s - \sum_{l \in A} \delta_{lrs} t_l - \sum_{l \in A} \delta_{lrs} \bar{c}_l$, where $(U_s - \sum_{l \in r} t_l)$ is the traveler's value of the trip and $\sum_{l \in r} \bar{c}_l$ is the operators' value of the trip.

In all the stochastic assignment game formulations in this section, $\alpha$ is a parameter that indicates how significant the perception variation is (since $\alpha^{-1}$ scales the deterministic component relative to a fixed scale of the random component). However, with travelers' and operators' valuations included in one equation, their perception of the same amount of monetary cost is not the same. Hence, we define different weights for travelers' and operators' costs: $\alpha_t$ and $\alpha_c$, respectively. The ratio of $\alpha_t$ to $\alpha_c$ represents the relative perception difference of travelers and operators. The absolute scale of $\alpha_t$ and $\alpha_c$ represents how significant the overall perception variation is, similar to the case with a single $\alpha$.

In deterministic assignment games, the price of a matching divides the characteristic function into two parts, which are buyer's payoff and seller's payoff without any loss. However, in the stochastic case, different perceptions of the same monetary cost lead to a loss of value. Denote the fare of link $l \in A$ as $p_l$. The traveler's perceived paid fare is $\alpha_t p_l$ while the operators' perceived received fare is $\alpha_c p_l$. The amount $(\alpha_t - \alpha_c) p_l$ is the loss associated with link $l \in A$, with the value referred to as a "loss" as we generally expect $\alpha_t \geq \alpha_c$. Further details on this point are discussed in section 2.4.1 (**Remark 1**). Hence, the actual characteristic function is $a_{rs} = \alpha_t (U_s - \sum_{l \in A} \delta_{lrs} t_l) - \alpha_c \sum_{l \in A} \delta_{lrs} \bar{c}_l - (\alpha_t - \alpha_c) \sum_{l \in A} \delta_{lrs} p_l$, which reduces to Eq. (22).

$$a_{rs} = \alpha_t U_s - \sum_{l \in A} \delta_{lrs}[\alpha_t(t_l + p_l) + \alpha_c(\bar{c}_l - p_l)], \quad \forall r \in R_s, s \in S \tag{22}$$

Similar to $P_3$, the decision variable $f_{rs}$ represents the flow of traveler group $s$ on path $r$. An entropy term is added to the objective function (Eq. 23a). To represent demand elasticity, we introduce a dummy link for each OD pair that directly connects the origin and destination. The cost of the dummy link corresponds to the perceived generalized cost of the outside option, and the flow on this link captures the portion of demand that is not served by the real network. With this addition, the demand constraint becomes an equality (Eq. 23c), and the model determines both the served demand on real links and the unserved portion assigned to the dummy link. Consequently, the first term $\alpha_t U_s$ of $a_{rs}$ becomes a constant and can be removed from the objective. The capacity constraints in Eq. (23b) now apply to the real links, and their inclusion here reflects that served demand is explicitly modeled in this formulation. The stochastic multimodal traffic assignment problem is then formulated as $P_5$, whose objective (Eq. 23a) is the sum of traveler and operator weighted costs and the entropy term. Traveler costs include both non-monetary costs and fares, while operator costs equal total operation cost minus total revenue.

$$P_5: \min_f \sum_{s \in S} \sum_{r \in R_s} f_{rs}(\ln(f_{rs}) - 1) + \sum_{s \in S} \sum_{r \in R_s} \sum_{l \in A} \delta_{lrs}[\alpha_t(t_l + p_l) + \alpha_c(\bar{c}_l - p_l)]f_{rs} \quad (23a)$$

Subject to

$$\sum_{s \in S} \sum_{r \in R_s} \delta_{lrs} f_{rs} \leq w_l, \quad \forall l \in A \quad (v_l) \quad (23b)$$

$$\sum_{r \in R_s} f_{rs} = d_s, \quad \forall s \in S \quad (u_s) \quad (23c)$$

$$f_{rs} \geq 0, \quad \forall r \in R_s, s \in S \quad (23d)$$

Similarly, we write the Lagrangian as Eq. (24) and take its derivative as Eq. (25).

$$L = \sum_{s \in S} \sum_{r \in R_s} f_{rs}(\ln(f_{rs}) - 1) + \sum_{s \in S} \sum_{r \in R_s} \sum_{l \in A} \delta_{lrs}[\alpha_t(t_l + p_l) + \alpha_c(\bar{c}_l - p_l)]f_{rs}$$
$$+ \sum_{l \in A} v_l (\sum_{s \in S} \sum_{r \in R_s} \delta_{lrs} f_{rs} - w_l) + \sum_{s \in S} u_s (\sum_{r \in R_s} f_{rs} - d_s) \quad (24)$$

$$\frac{\partial L}{\partial f_{rs}} = \ln(f_{rs}) + \delta_{lrs}[\alpha_t(t_l + p_l) + \alpha_c(\bar{c}_l - p_l)] + \sum_{l \in A} \delta_{lrs} v_i + u_s \quad (25)$$

Eq. (25) reduces to Eq. (26) with the KKT conditions, which has the same kind of structure as Eq. (17). The right-hand-side consists the negative of total costs of travelers and operators, the negative of traveler's and operators' expected payoffs, in which $u_s$ is the expected payoff of a traveler of group $s \in S$, $\sum_{l \in A} \delta_{lrs} v_i$ is operators' expected payoff (also link delays), $\delta_{lrs}[\alpha_t(t_l + p_l) + \alpha_c(\bar{c}_l - p_l)]$ is the cost of the matching.

$$\ln(f_{rs}) = -\delta_{lrs}[\alpha_t(t_l + p_l) + \alpha_c(\bar{c}_l - p_l)] - \sum_{l \in A} \delta_{lrs} v_i - u_s, \quad \forall i \in Q, j \in P \quad (26)$$

### 2.4 Stochastic assignment game in urban mobility markets: stochastic multimodal assignment game

#### 2.4.1 Formulation of the stochastic multimodal assignment game

We change the "sellers" from links to the operators in urban mobility markets and derive the **stochastic multimodal assignment game**. Each operator can own multiple links/nodes in the network, and they decide which links they operate, and what capacity/fleet size they choose to provide. We still use $Q$ to denote the set of operators, which include both fixed-route transit and MoD operators. Travelers and operators make joint decisions. Operators decide where to operate, and what capacity/fleet size to provide. Travelers choose paths. We construct a network structure that connects walking subnetworks, fixed-route subnetworks, and MoD subnetworks similar to Liu and Chow (2024). All the costs are assumed to be in a common transferable unit (e.g. $).

Consider a platform defined as a directed graph $G(N, A)$ serving a set of user OD pairs $S$ traveling to/from centroids $N_Z \subset N$. Each link in $A$ represents a service route between two nodes (e.g. centroid to fixed route station, station to station, centroid to centroid, or wait time to transfer from one mode to another). Each OD pair $s \in S$ consists of a population of travelers $d_s$. Travelers' decisions are modeled by stochastic path flows $f_{rs}$ for path $r \in R_s, \forall s \in S$, where $R_s$ is a predefined path set connecting OD pair $s \in S$. Parameter $\delta_{lrs} \in \{0,1\}$ denotes whether link $l \in A$ is on path $r \in R_s$. Dummy links $\ell_s$ are added for the OD pair of all traveler groups $s \in S$ to capture demand elasticity. Each dummy link directly connects the origin and destination nodes of OD pair $s$ and represents travelers who opt out of the platform. The cost of the dummy link is set as a constant disutility calibrated to reflect the perceived utility of external travel alternatives (e.g., competing platforms, private vehicles, or no trip). The dumpy link has zero operating cost and infinite capacity.

There exists a set of fixed-route operators $Q_F$ such that each operator $f \in Q_F$ deterministically owns one or more service links, $A_f \subseteq A_F$. Each link $l$ is associated with an operation cost $c_l$ given a maximum service capacity that can be provided $w_l$, and a travel cost $t_l$ (which includes all non-monetary costs excluding fares, such as in-vehicle travel time, average wait time, comfort cost, etc.). Unit cost of operation on link $l$ can be computed as $c_l/w_l$.

A set of MoD operators $f \in Q_M$ considers operating candidate nodes $N_M \subset N$ in this platform. $N_M$ is the union of candidate nodes sets $N_f$ of all MoD operators $f \in Q_M$. These nodes represent service zones. Each MoD zone $i \in N_M$ is associated with an operation cost $q_i$ given a maximum fleet size $z_i$ that can be deployed. Unit cost of operation on link $l$ can be computed as $q_i/z_i$. For each operator $f \in Q_M$, the nodes $N_f$ are connected as complete subgraphs to each other via uncapacitated MoD links $A_f = N_f \times N_f$. The set of all MoD links is denoted as $A_M = \bigcup_{f \in Q_M} A_f$. The set of all the outbound links from node $i \in N_M$ is denoted $A_i^-$. Operation cost per flow unit between MoD nodes connected by link $l \in A_M$ is defined as $m_l$. Transfer links $A_{0M}$ connect the centroids $N_Z$ to the MoD nodes $N_M$, including access links (transfer to MoD) and egress links (transfer from MoD). $A_{0f}$ denotes access links to the services provided by operator $f \in Q_M$. Travel costs of transfer links are uncongested time costs of accessing/egressing. The network structure is illustrated in **Figure 1**.

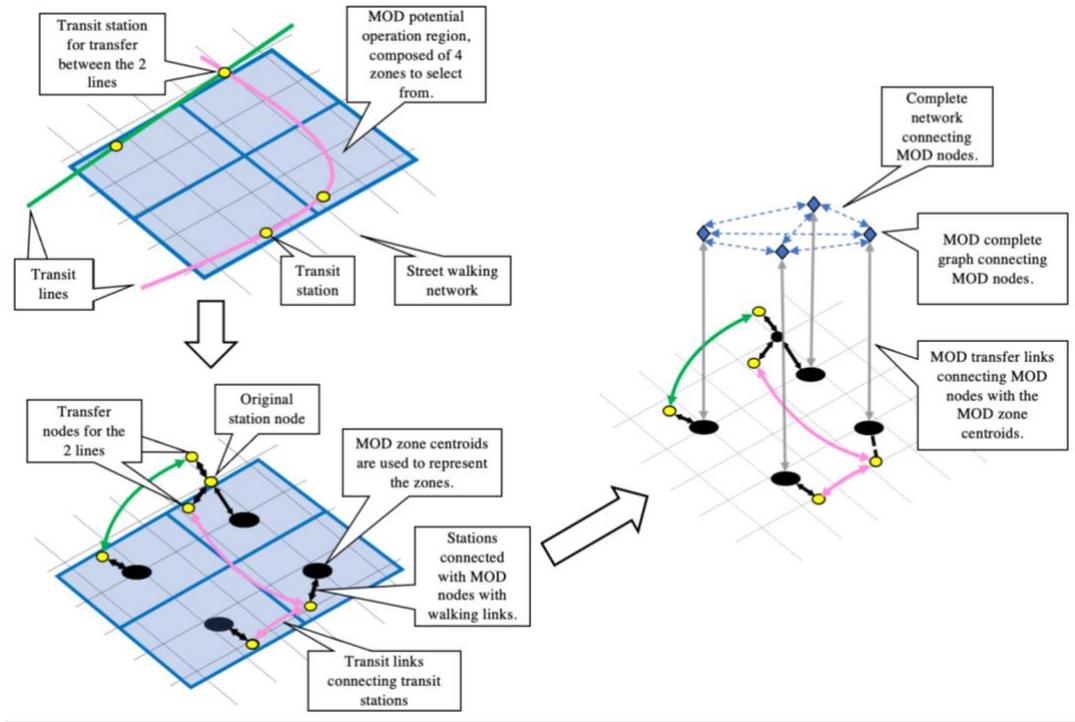

**Figure 1.** Illustrating network structure.

One difference between the stochastic multimodal assignment game and the deterministic version proposed by Liu and Chow (2024) is that the decisions of operators become probabilistic. Instead of modeling operators' decision with binary variables, we use continuous variables between 0 and 1 to find the probabilities of a link/node being operated. Variable $y_l$ ($0 < y_l \leq 1$) is the probability that a fixed-route transit link $l$ opens. Variable $v_i$ ($0 < v_i \leq 1$) is the probability that a MoD zone $i$ opens. Both variables are non-zero. With a path-based logit formulation, all the links and nodes (zones) covered by the predefined path set have non-zero flows, which lead to non-zero operating probabilities. With the maximum link capacity $w_l$ and maximum fleet size $z_i$, expected link capacity of link $l$ and expected fleet size of zone $i$ can be computed accordingly: $w_l y_l$ and $z_i v_i$, respectively.

As deduced in section 2.3, with different perceptions of the same monetary cost, the weights of travelers' and operators' valuations in the characteristic function are different, leading to a loss in the transactions of value. Link fares become parameters that need to be known to solve the stochastic multimodal assignment game. We denote the preset link fare of link $l \in A_F \cup A_M$ as $\widetilde{p}_l$. Travelers and operators make joint decisions with the fares pre-set. The stochastic multimodal assignment game can be formulated as $P_6$.

$$P_6: \min \Phi_1 = \sum_{s \in S} \sum_{r \in R_s} f_{rs} (\ln(f_{rs}) - 1)$$

$$+ \alpha_t \left( \sum_{s \in S} \sum_{r \in R_s} \sum_{l \in A_F \cup A_M \cup A_D \cup A_{0M}} t_l \delta_{lrs} f_{rs} + \sum_{f \in Q} \sum_{l \in A_f} \tilde{p}_l \sum_{s \in S} \sum_{r \in R_s} \delta_{lrs} f_{rs} \right)$$

$$+ \alpha_c \left( \sum_{s \in S} \sum_{r \in R_s} \sum_{l \in A_M} m_l \delta_{lrs} f_{rs} + \sum_{l \in A_F} c_l y_l + \sum_{i \in N_M} q_i v_i \right.$$

$$\left. - \sum_{f \in Q} \sum_{l \in A_f} \tilde{p}_l \sum_{s \in S} \sum_{r \in R_s} \delta_{lrs} f_{rs} \right) \quad (27a)$$

Subject to

$$\sum_{r \in R_s} f_{rs} = d_s, \forall s \in S \quad (u_s) \quad (27b)$$

$$\sum_{s \in S} \sum_{r \in R_s} \delta_{lrs} f_{rs} \leq w_l y_l, \forall l \in A_F \quad (\mu_l) \quad (27c)$$

$$\sum_{l \in A_i^-} \sum_{s \in S} \sum_{r \in R_s} \delta_{lrs} f_{rs} \leq z_i v_i, \forall i \in N_M \quad (\lambda_i) \quad (27d)$$

$$y_l \leq 1, \quad \forall l \in A_F \quad (\gamma_l) \quad (27e)$$

$$v_i \leq 1, \quad \forall i \in N_M \quad (\pi_i) \quad (27f)$$

$$f_{rs} \geq 0, \quad \forall r \in R_s, s \in S \quad (27g)$$

$$y_l \geq 0, \quad \forall l \in A_F \quad (27h)$$

$$v_i \geq 0, \quad \forall i \in N_M \quad (27i)$$

$P_6$ has the same structure as $P_5$. The objective function Eq. (27a) is the sum of weighted costs of travelers and operators and the entropy term. We name the sum of weighted costs of travelers and operators as the **system cost**. Constraints Eq. (27b) is the demand constraint. Constraint Eq. (27c) is the capacity constraint for fixed-route links. Constraints (27d) are the MoD zone fleet size constraints for MoD nodes corresponding to MoD zones, where the left-hand-side (LHS) is the total outbound flow of MoD node $i \in N_M$. The corresponding Lagrange multipliers are labeled in parentheses.

Assumptions are as follows:

- We assume that the cost of fixed-route transit links and MoD nodes are linear w.r.t capacities and fleet sizes, hence we can compute the expected operation costs of fixed-route link $l$ and MoD zone $i$: $c_l y_l / w_l$ and $q_i v_i / z_i$, respectively.

Unlike the traditional stochastic user equilibrium (SUE) (Bell, 1995), which only considers users' heterogeneity, stochastic path flows solved from the stochastic multimodal assignment game incorporate heterogeneity of both users and operators, which are interdependent and impossible to separate. In **Proposition 1**, we show that $P_6$ yields path flows that conforms to a coalitional logit model the decision-maker is the coalition formed by the travelers matching with the operators. The disutility function of path $r \in R_s, s \in S$ is the sum of a traveler's total cost weighted by $\alpha_t$ and operators' total cost weighted by $\alpha_c$.

**Proposition 1.** Logit path flows with coalitional decision-making. *$P_6$ yields path flows that conform to a logit model where the disutility function of path $r \in R_s, s \in S$ is the sum of a user's total cost of path $r$ weighted by $\alpha_t$ and operators' total cost of transporting one user on path $r$ weighted by $\alpha_c$.*

***Proof.*** The Lagrangian equation of $P_6$ is written as Eq. (28). Its derivative is written as Eq. (29). For the KKT conditions, we have $\frac{\partial L}{\partial f_{rs}} = 0$ at optimum.

$$\begin{aligned}
L = & \sum_{s \in S} \sum_{r \in R_s} f_{rs} (\ln(f_{rs}) - 1) \\
& + \alpha_t \left( \sum_{s \in S} \sum_{r \in R_s} \sum_{l \in A_F \cup A_M \cup A_D \cup A_{0M}} t_l \delta_{lrs} f_{rs} + \sum_{f \in Q} \sum_{l \in A_f} \tilde{p}_l \sum_{s \in S} \sum_{r \in R_s} \delta_{lrs} f_{rs} \right) \\
& + \alpha_c \left( \sum_{s \in S} \sum_{r \in R_s} \sum_{l \in A_M} m_l \delta_{lrs} f_{rs} + \sum_{l \in A_F} c_l y_l + \sum_{i \in N_M} q_i v_i \right) \\
& - \sum_{f \in Q} \sum_{l \in A_f} \tilde{p}_l \sum_{s \in S} \sum_{r \in R_s} \delta_{lrs} f_{rs} - \sum_{s \in S} u_s \left( \sum_{r \in R_s} f_{rs} - d_s \right) \\
& - \sum_{l \in A_F} \mu_l \left( \sum_{s \in S} \sum_{r \in R_s} \delta_{lrs} f_{rs} - w_l y_l \right) \\
& - \sum_{i \in N_M} \lambda_i \left( \sum_{l \in A_i^-} \sum_{s \in S} \sum_{r \in R_s} \delta_{lrs} f_{rs} - z_i v_i \right) - \sum_{l \in A_F} \gamma_l (y_l - 1) \\
& - \sum_{i \in N_M} \pi_i (v_i - 1)
\end{aligned} \quad (28)$$

$$\frac{\partial L}{\partial f_{rs}} = \ln(f_{rs}) + \alpha_t \left( \sum_{l \in A_F \cup A_M \cup A_D \cup A_{0M}} t_l \delta_{lrs} + \sum_{f \in Q} \sum_{l \in A_f} \widetilde{p}_l \delta_{lrs} \right)$$
$$+ \alpha_c \left( \sum_{l \in A_M} m_l \delta_{lrs} - \sum_{f \in Q} \sum_{l \in A_f} \widetilde{p}_l \delta_{lrs} \right) - u_s - \sum_{l \in A_F} \mu_l \delta_{lrs} \quad (29)$$
$$- \sum_{i \in N_M} \lambda_i \sum_{l \in A_i^-} \delta_{lrs} = 0$$

Since the path flows are strictly positive, all the $y_l$ and $v_i$ for the links/zones that are covered by the path set are strictly positive. Hence, the optimality conditions of $y_l$ and $v_i$ can be written as Eq. (30) and Eq. (31), from which we can solve for $\mu_l$ and $\lambda_i$ as shown in Eq. (32) and Eq. (33).

$$\frac{\partial L}{\partial y_l} = \alpha_c c_l - \mu_l w_l + \gamma_l = 0 \quad (30)$$

$$\frac{\partial L}{\partial v_i} = \alpha_c q_i - \lambda_i z_i + \pi_i = 0 \quad (31)$$

$$\mu_l = \frac{\alpha_c c_l + \gamma_l}{w_l} \quad (32)$$

$$\lambda_i = \frac{\alpha_c q_i + \pi_i}{z_i} \quad (33)$$

Substituting Eqs. (32) – (33) into Eq. (29) leads to Eq. (34).

$$\ln(f_{rs}) = -\alpha_t \left( \sum_{l \in A_F \cup A_M \cup A_D \cup A_{0M}} t_l \delta_{lrs} - \sum_{l \in A_F} \frac{\gamma_l}{w_l \alpha_t} \delta_{lrs} - \sum_{i \in N_M} \frac{\pi_i}{z_i \alpha_t} \sum_{l \in A_i^-} \delta_{lrs} \right.$$
$$\left. + \sum_{f \in Q} \sum_{l \in A_f} \widetilde{p}_l \delta_{lrs} \right) \quad (34)$$
$$- \alpha_c \left( \sum_{l \in A_M} m_l \delta_{lrs} + \sum_{l \in A_F} \frac{c_l}{w_l} \delta_{lrs} + \sum_{i \in N_M} \frac{q_i}{z_i} \sum_{l \in A_i^-} \delta_{lrs} - \sum_{f \in Q} \sum_{l \in A_f} \widetilde{p}_l \delta_{lrs} \right)$$
$$- u_s$$

We show that $\frac{-\gamma_l}{w_l \alpha_t}$ is equivalent to the link delay of fixed-route link $l \in A_F$, and $\frac{-\pi_i}{z_i \alpha_t}$ is equivalent to the link delay of all outbound links of MoD node $i \in N_M$ in **Proposition 2**. If we denote the delays as $D_l^{A_F} = \frac{-\gamma_l}{w_l \alpha_t}$ for $l \in A_F$, and $D_i^{N_M} = \frac{-\pi_i}{z_i \alpha_t}$ for $i \in N_M$, Eq. (34) can be rewritten as Eq. (35).

$$\ln(f_{rs}) = -\alpha_t \left( \sum_{l \in A_F \cup A_M \cup A_D \cup A_{0M}} t_l \delta_{lrs} + \sum_{l \in A_F} D_l^{A_F} \delta_{lrs} + \sum_{i \in N_M} D_i^{N_M} \sum_{l \in A_j^-} \delta_{lrs} \right.$$
$$\left. + \sum_{f \in Q} \sum_{l \in A_f} \tilde{p}_l \delta_{lrs} \right) \tag{35}$$

$$- \alpha_c \left( \sum_{l \in A_M} m_l \delta_{lrs} + \sum_{l \in A_F} \frac{c_l}{w_l} \delta_{lrs} + \sum_{i \in N_M} \frac{q_i}{z_i} \sum_{l \in A_i^-} \delta_{lrs} - \sum_{f \in Q} \sum_{l \in A_f} \tilde{p}_l \delta_{lrs} \right)$$
$$- u_s$$

We define $T_{rs}$ and $C_{rs}$ in Eqs. (36) – (37).

$$T_{rs} = \sum_{l \in A_F \cup A_M \cup A_D} t_l \delta_{lrs} + \sum_{l \in A_F} D_l^{A_F} \delta_{lrs} + \sum_{i \in N_M} D_i^{N_M} \sum_{l \in A_j^-} \delta_{lrs} + \sum_{f \in Q} \sum_{l \in A_f} \tilde{p}_l \delta_{lrs} \tag{36}$$

$$C_{rs} = \sum_{l \in A_{MOD}} m_l \delta_{lrs} + \sum_{l \in A_F} \frac{c_l}{w_l} \delta_{lrs} + \sum_{i \in N_M} \frac{q_i}{z_i} \sum_{l \in A_i^-} \delta_{lrs} - \sum_{f \in Q} \sum_{l \in A_f} \tilde{p}_l \delta_{lrs} \tag{37}$$

Eq. (35) can be rewritten as Eq. (38), leading to the logit flow shown in Eq. (39). The disutility function can be written as Eq. (40). Eq. (38) is consistent with Eq. (26), in which the right-hand-side consists of the negative of traveler and operators total expected payoffs and the negative of total costs of matching. The difference is that, in Eq. (38), operators' expected payoff is transformed into link/zone delays ($D_l^{A_F}, D_i^{N_M}$) and added into $T_{rs}$, whereas $u_s$ is still the expected payoff of a traveler of group $s \in S$.

$$\ln(f_{rs}) = -\alpha_t T_{rs} - \alpha_c C_{rs} - u_s \tag{38}$$

$$f_{rs} = d_s \frac{\exp(-\alpha_t T_{rs} - \alpha_c C_{rs})}{\sum_{r' \in R_s} \exp(-\alpha_t T_{r's} - \alpha_c C_{r's})} \tag{39}$$

$$u_r = -\alpha_t T_{rs} - \alpha_c C_{rs}, \qquad \forall r \in R_s, s \in S \tag{40}$$

The disutility function Eq. (40) is a combination of users' disutility and operators' disutility. $T_{rs}$ is the travel cost of one user on path $r$, which is the sum of travel costs of links (delay included) and total fare paid. $C_{rs}$ is the operators' cost of operating path $r$ for one user—including unit cost of MoD links (first term), fixed-route links (second term), and deploying fleets in MoD service zones represented by MoD nodes (third term)—minus the total fare of one user (fourth term). ∎

Eq. (39) is a coalitional logit model of users and operators. In a deterministic many-to-many assignment game, the fares are solved given the cost allocation to ensure the stability of matchings solved from the matching problem, instead of directly including the impact of fares in the matching problem. This is equivalent to the situation that, when $\alpha_t = \alpha_c$, fares become a value transferred from users to operators completely, which cancels out. When $\alpha_t \neq \alpha_t$, the role of fares is captured. The relative magnitude of $\alpha_t$ and $\alpha_t$ represents users' and operators' different sensitivity of the same amount of monetary cost. The absolute magnitudes of $\alpha_t$ and $\alpha_t$ influence how dispersed the flows are. The smaller $\alpha_t$ and $\alpha_t$ are, the more diverse users and operators are, and therefore the more spread-out the path flows are.

**Remark 1**: *Holding other factors constant, path flows decrease when fares increase. It can be inferred that users are more sensitive than operators considering the same amount of monetary cost ($\alpha_t > \alpha_c$).*

**Proposition 2** gives the delay terms of fixed-route links and outbound links of MoD nodes.

**Proposition 2.** Delays caused by capacities and fleet sizes. *At the equilibrium of the stochastic many-to-many assignment game, link delay of fixed-route link $l \in A_F$ equals to $\frac{-\gamma_l}{w_l \alpha_t}$, link delay of outbound links of MoD node $i \in N_M$ equals to $\frac{-\pi_j}{z_j \alpha_t}$.*

*Proof*: 1) Fixed-route link delay:

Denote flow on link $l \in A_F$ as $h_l$. When $h_l = w_l$, if $\delta$ units of flow is added to link $l$, $h_l = w_l + \delta$. It can be computed that $y_l = \frac{w_l + \delta}{w_l}$, which exceeds the right-hand-side of constraint Eq.(27e) by $\frac{\delta}{w_l}$. Assume the original path of the $\delta$ units is path $r$, and the path which the $\delta$ units switched to is path $r'$. We denote the path flow, travel cost, fare, and operation cost per user of path $r$ as $f_r, t_r, \tilde{p}_r$, and $c_r$, and path $r'$ parameters similarly as $t_{r'}, \tilde{p}_{r'}$, and $c_{r'}$. The change of the objective value of $P_6$ is written as Eq. (41).

$$d\Phi_1 = \delta \ln\left(\frac{f_r}{f_{r'}}\right) + \delta\alpha_t(t_r - t_{r'}) + \delta\alpha_c(c_r - c_{r'}) + \delta(\alpha_t - \alpha_c)(\widetilde{p_r} - \widetilde{p_{r'}}) \qquad (41)$$

Consider the capacity constraint Eq. (27e). Increasing the available capacity of link $l$ by oen unit increases the objective value of $P_6$ by its associated dual variable $\gamma_l$. Hence, the marginal effect of adding one unit of capacity on link $l$ leads to Eq. (42). Eq. (42) can then be rewritten as Eq. (43).

$$\frac{d\Phi_1}{\delta/w_l} = w_l \left(\ln\left(\frac{f_r}{f_{r'}}\right) + \alpha_t(t_r - t_{r'}) + \alpha_c(c_r - c_{r'}) + (\alpha_t - \alpha_c)(\widetilde{p_r} - \widetilde{p_{r'}})\right) = \gamma_l \qquad (42)$$

$$\ln\left(\frac{f_r}{f_{r'}}\right) = -\alpha_t\left(t_r - \frac{\gamma_l}{w_l \alpha_t} - t_{r'}\right) - \alpha_c(c_r - c_{r'}) - (\alpha_t - \alpha_c)(\widetilde{p_r} - \widetilde{p_{r'}}) \qquad (43)$$

The coalitional logit model can be written as Eq. (44).

$$\ln\left(\frac{f_r}{f_{r'}}\right) = -\alpha_t(t_r + D_l^{A_F} - t_{r'}) - \alpha_c(c_r - c_{r'}) - (\alpha_t - \alpha_c)(\widetilde{p_r} - \widetilde{p_{r'}}) \tag{44}$$

Comparing Eq. (43) and Eq. (44), we have Eq. (45), which shows link delay of link $l \in A_F$ when $v_l = w_l$.

$$D_l^{A_F} = -\frac{\gamma_l}{w_l \alpha_t} \tag{45}$$

When $h_l < w_l$, $y_l = \frac{v_l}{w_l} < 1$. According to complimentary slackness, $\gamma_l = 0$, and $D_l^{A_F} = -\frac{\gamma_l}{w_l \alpha_t} = 0$.

*2) MoD node outbound link delay:*

Denote total flow going through MoD node $i \in N_M$ as $h_i$. The logic is the same as the link delays of fixed-route links. When $h_i = z_i$, if $\delta$ units of flow is added to a outbound link $l$ of MoD node $i$, $h_i = z_i + \delta$. It can be computed that $v_i = \frac{z_i + \delta}{z_i}$, which exceeds the right-hand-side of constraint (27f) by $\frac{\delta}{z_i}$. Assume the original path of the $\delta$ units is path $r$, and the path which the $\delta$ units switched to is path $r'$. The change of the objective value of $P_6$ is written as Eq. (41).

When the allowable bound in Eq. (27f) is increased by one unit, the objective value of $P_6$ increases by its associated dual variable $\pi_i$. Hence, the marginal effect of this change leads to Eq. (46), which can then be rewritten as Eq. (47).

$$\frac{d\Phi_1}{\delta/z_i} = z_i \left( \ln\left(\frac{f_r}{f_{r'}}\right) + \alpha_t(t_r - t_{r'}) + \alpha_c(c_r - c_{r'}) + (\alpha_t - \alpha_c)(\widetilde{p_r} - \widetilde{p_{r'}}) \right) = \pi_i \tag{46}$$

$$\ln\left(\frac{f_r}{f_{r'}}\right) = -\alpha_t\left(t_r - \frac{\pi_i}{z_i \alpha_t} - t_{r'}\right) - \alpha_c(c_r - c_{r'}) - (\alpha_t - \alpha_c)(\widetilde{p_r} - \widetilde{p_{r'}}) \tag{47}$$

The coalitional logit model can be written as Eq. (48).

$$\ln\left(\frac{f_r}{f_{r'}}\right) = -\alpha_t(t_r + D_i^{N_M} - t_{r'}) - \alpha_c(c_r - c_{r'}) - (\alpha_t - \alpha_c)(\widetilde{p_r} - \widetilde{p_{r'}}) \tag{48}$$

Comparing Eq. (47) and Eq. (48), we have Eq. (49), which shows link delay of outbound link $l$ from MOD node $i \in N_M$ when $v_l = w_l$.

$$D_i^{N_M} = -\frac{\pi_i}{z_i \alpha_t} \tag{49}$$

When $h_i < z_i$, $v_i = \frac{v_l}{z_i} < 1$. According to complimentary slackness, $\pi_i = 0$, and $D_i^{N_M} = -\frac{\pi_i}{z_i \alpha_t} = 0$. ∎

### 2.4.2 Solution algorithm to the stochastic multimodal assignment game

We adapt the iterative balancing algorithm from (Bell, 1995) to solve $P_6$. Stochastic path flow from $P_6$ can be written as Eq. (50).

$$f_{rs} = \exp\left(-\alpha_t \left(\sum_{l \in A_F \cup A_M \cup A_D \cup A_{0M}} t_l \delta_{lrs} + \sum_{f \in Q} \sum_{l \in A_f} \tilde{p}_l \delta_{lrs}\right)\right.$$
$$- \alpha_c \left(\sum_{l \in A_M} m_l \delta_{lrs} + \sum_{l \in A_F} \frac{c_l}{w_l} \delta_{lrs} + \sum_{i \in N_M} \frac{q_i}{z_i} \sum_{l \in A_i^-} \delta_{lrs}\right) \quad (50)$$
$$\left.- \sum_{f \in Q} \sum_{l \in A_f} \tilde{p}_l \delta_{lrs}\right) \exp(\beta_s) \prod_{l \in A_F} \exp\left(\frac{\gamma_l}{w_l} \delta_{lrs}\right) \prod_{j \in N_M} \prod_{l \in A_j^-} \exp\left(\frac{\pi_j}{z_j} \delta_{lrs}\right)$$

If we define $B_s = \exp(\beta_s)$, $M_l^{A_F} = \exp\left(\frac{\gamma_l}{w_l}\right)$, and $M_l^{N_M} = \exp\left(\frac{\pi_j}{z_j}\right)$, Eq. (45) can be written as Eq. (51).

$$x_{rs} = \exp\left(-\alpha_t \left(\sum_{l \in A_F \cup A_M \cup A_D \cup A_{0M}} t_l \delta_{lrs} + \sum_{f \in Q} \sum_{l \in A_f} \tilde{p}_l \delta_{lrs}\right)\right.$$
$$- \alpha_c \left(\sum_{l \in A_M} m_l \delta_{lrs} + \sum_{l \in A_F} \frac{c_l}{w_l} \delta_{lrs} + \sum_{i \in N_M} \frac{q_i}{z_i} \sum_{l \in A_i^-} \delta_{lrs}\right) \quad (51)$$
$$\left.- \sum_{f \in Q} \sum_{l \in A_f} \tilde{p}_l \delta_{lrs}\right) B_s \prod_{l \in A_F, l \in r} M_l^{A_F} \prod_{j \in N_M} \prod_{l \in A_j^-, l \in r} M_l^{N_M}$$

**Algorithm 1** finds the values of $B_s$, $M_l^{A_F}$, and $M_l^{N_M}$ satisfying the optimality condition (Eq. (35)). **Proposition 3** establishes the convergence guarantee.

**Algorithm 1.** Iterative Balancing.

**Inputs:** network parameters: $t, m, c, w, q, z,$; demand: $d$; weights: $\alpha_t, \alpha_c$; path set: $R_s, s \in S$; current fares $\tilde{p}$.

**Step 1** (initialize):
  $M_l^{A_F} = 1$ for all $l \in A_F$
  $M_l^{N_M} = 1$ for all $i \in N_M$
  $B_s = 1$ for all $s \in S$

**Step 2** (iterate):
  Repeat the following until convergence
    Compute all path flows $\hat{f}_{rs}, r \in R_s, s \in S$ with Eq. (34).
      For each $l \in A_F$ compute

$$\theta = \frac{w_l}{\sum_{s \in S} \sum_{r \in R_s} \delta_{lrs} \hat{f}_{rs}}$$

$$M_l^{A_F} = \min(1, \theta M_l^{A_F})$$

For each $l \in A_i^-, i \in N_M$ compute

$$\theta = \frac{z_i}{\sum_{l \in A_i^-} \sum_{s \in S} \sum_{r \in R_s} \delta_{lrs} \hat{f}_{rs}}$$

$$M_l^{N_M} = \min(1, \theta M_l^{N_M})$$

For each $s \in S$ compute

$$\theta = \frac{d_s}{\sum_{r \in R_s} \hat{f}_{rs}}$$

$$B_s = \theta B_s$$

**Step 3** (compute optimal path flows and delays):

Compute optimal path flows $\tilde{f}_{rs}, r \in R_s, s \in S$ using Eq. (34).

For each $l \in A_F$ compute

$$\tilde{y}_l = \frac{\sum_{s \in S} \sum_{r \in R_s} \delta_{lrs} \tilde{f}_{rs}}{w_l}$$

$$D_l^{A_F} = -\ln(M_l^{A_F})/\alpha_t$$

For each $l \in A_i^-, i \in N_M$ compute

$$\tilde{v}_i = \frac{\sum_{l \in A_i^-} \sum_{s \in S} \sum_{r \in R_s} \delta_{lrs} \tilde{f}_{rs}}{z_i}$$

$$D_i^{N_M} = -\ln(M_l^{N_M})/\alpha_t$$

**Outputs:** $\tilde{f}, \tilde{y}, \tilde{v}, D^{A_F}, D^{N_M}$

**Proposition 3.** Convergence of iterative balancing. *Iterative balancing algorithm (Algorithm 1) converges to the solution of $P_6$ if a feasible solution exists.*

**Proof.** Decision variables of operators can be computed from the flows as Eqs. (52) – (53).

$$y_l = \frac{\sum_{s \in S} \sum_{r \in R_s} \delta_{lrs} f_{rs}}{w_l} \quad (52)$$

$$v_i = \frac{\sum_{l \in A_i^-} \sum_{s \in S} \sum_{r \in R_s} \delta_{lrs} f_{rs}}{z_i} \quad (53)$$

Plugging Eqs. (32), (33), (52), (53) into the Lagrangian, we obtain Eq. (54).

$$L = \sum_{s \in S} \sum_{r \in R_s} f_{rs} \left( \ln(f_{rs}) - 1 \right)$$

$$+ \alpha_t \left( \sum_{s \in S} \sum_{r \in R_s} \sum_{l \in A_F \cup A_M \cup A_D \cup A_{0M}} t_l \delta_{lrs} f_{rs} + \sum_{f \in Q} \sum_{l \in A_{0f}} \tilde{p}_l \sum_{s \in S} \sum_{r \in R_s} \delta_{lrs} f_{rs} \right)$$

$$+ \alpha_c \left( \sum_{s \in S} \sum_{r \in R_s} \sum_{l \in A_M} m_l \delta_{lrs} f_{rs} - \sum_{f \in Q} \sum_{l \in A_{0f}} \tilde{p}_l \sum_{s \in S} \sum_{r \in R_s} \delta_{lrs} f_{rs} \right)$$

$$+ \sum_{s \in S} \beta_s \left( \sum_{r \in R_s} f_{rs} - d_s \right) \tag{54}$$

$$+ \sum_{l \in A_F} \left( \left( \frac{\alpha_c c_l}{w_l} + \ln(M_l^{A_F}) \right) \sum_{s \in S} \sum_{r \in R_s} \delta_{lrs} f_{rs} - 1 \right)$$

$$+ \sum_{i \in N_M} \sum_{l \in A_i^-} \left( \left( \frac{\alpha_c q_i}{z_i} + \ln(M_l^{N_M}) \right) \sum_{s \in S} \sum_{r \in R_s} \delta_{lrs} f_{rs} - 1 \right)$$

At the optimum, $L$ is minimized with respect to the primal variables ($f_{rs}$, $y_l$, $v_i$) and maximized with respect to the dual variables ($\gamma_l$, $\pi_i$).

First, we consider the capacity constraints of fixed-route links. Denote flow on link $l$ as $v_l$. If $h_l > w_l$ for a fixed link $l \in A_F$, $L$ is increased by reducing $M_l^{A_F}$, which increases $D_l^{A_F}$. This also reduces $v_l$, so $M_l^{A_F}$ should be reduced until $h_l = w_l$. If $h_l < w_l$, $L$ is increased by increasing $M_l^{A_F}$ until $h_l = w_l$ or $M_l^{A_F} = 1$ ($D_l^{A_F} = 0$).

Then, we consider the MOD node capacity constraints. Denote total flow going through MOD node $i$ as $h_i$. If $h_i > z_i$ for a MOD node $i \in N_M$, $L$ is increased by reducing $M_l^{N_M}$, which increases $D_l^{N_M}$. This reduces $v_i$, so $M_l^{N_M}$ should be reduced until $h_i = z_i$. If $h_i < z_i$, $L$ is increased by increasing $M_l^{N_M}$ until $h_i = z_i$ or $M_l^{N_M} = 1$ ($D_l^{N_M} = 0$).

Finally, we consider the demand constraints. Denote total flow assigned for an OD pair as $h_s$. If $h_s > d_s$, $L$ is increased by reducing $B_s$ until $h_s = d_s$. If $h_s < d_s$, $L$ is increased by increasing $B_s$ until $h_s = d_s$.

The above describes Step 2 of the **Algorithm 1**. Each iteration leads to an increase in $L$. Iterations continue until the optimality conditions are satisfied. ∎

## 2.5 Integrate the role of the regulator or the MaaS platform: the stochastic multimodal Stackelberg game

The stochastic multimodal assignment game represents a systematic perspective, i.e. the perspective of a regulator or of a MaaS platform. In the deterministic multimodal assignment game studied by Liu and Chow (2024), the regulator can set fares and subsidies to achieve system goals, which can be system cost, emissions, equity measures, etc., and whether it is seller-optimal or buyer-optimal outcomes. In deterministic cases, one system equilibrium can be stabilized by a space of sets of fares and subsidies. However, in the stochastic multimodal assignment game ($P_6$), fares impact the choices of travelers and operators in a more sensitive way: one equilibrium is achieved by one set of fares. This difference between deterministic and stochastic systems is also observed in day-to-day processes (see Smith et al., 2014). The regulator (or the MaaS platform) impacts the decisions of travelers and operators through fare setting. This can be interpreted as a Stackelberg game where the regulator is the leader, and travelers and operators are the followers. The regulator fully anticipates how the travelers and operators react to the fare settings, while travelers and operators do not have information regarding how the regulator sets the fares.

Such a problem can be modeled through a bilevel problem: an upper-level problem of fare design from the regulator's perspective (leader), and a lower-level assignment game between operators and users (followers) which is $P_6$. Decision variables include path flows $f_{rs}$, operation decisions $y_l$ and $v_i$, and fares. Fare of link $l \in A_F \cup A_M$ set by the platform is modeled by decision variable $p_l, l \in A_f \cup A_M$. The resulting stochastic multimodal Stackelberg game $P_7$ is shown as Eq. (55). We define the solution of $P_7$ as a ***stochastic assignment game Stackelberg equilibrium***.

$$P_7: \max \Phi_2 = \sum_{s \in S} \sum_{r \in R_s} f_{rs} \sum_{l \in A_r} p_l \quad (55a)$$

subject to Eqs. (27b) – (27d) and

$$\sum_{l \in A_f} \sum_{s \in S} \sum_{r \in R_s} \delta_{lrs} f_{rs} p_l \geq \sum_{l \in A_f} c_l y_l, \quad \forall f \in Q_F \quad (55b)$$

$$\sum_{l \in A_f} p_l \sum_{s \in S} \sum_{r \in R_s} \delta_{lrs} f_{rs} \geq \sum_{l \in A_f} \sum_{s \in S} m_l \sum_{r \in R_s} \delta_{lrs} f_{rs} + \sum_{i \in N_f} q_i v_i, \quad \forall f \in Q_M \quad (55c)$$

$$\boldsymbol{f}, \boldsymbol{y}, \boldsymbol{v} = \arg\min \Phi_1(\boldsymbol{p}) \quad (55d)$$

$$f_{rs} \geq 0, \quad \forall r \in R_s, s \in S \quad (55e)$$

$$p_l \geq 0, \quad \forall l \in A_r, r \in R_s, s \in S \quad (55f)$$

$$0 \leq y_l \leq 1, \quad \forall l \in A_F \quad (55g)$$

$$0 \leq v_i \leq 1, \quad \forall i \in N_M \quad (55h)$$

Objective function Eq. (55a) finds the maximum revenue, which may be substituted with a different objective if desired (e.g. welfare or equity maximizing objective for publicly owned platforms). Constraint Eq. (55b) and Eq. (55c) are the profitability constraints of fixed-route operators and MoD operators, making sure that each operator's revenue covers the operator cost. Constraint Eq. (55d) is the equilibrium condition formulated into a lower-level problem $P_6$, representing how the stochastic path flows ($f$) and decisions of operators ($y, v$) react to the fare settings ($p$). Constraints Eqs. (55e) – (55h) are non-negativity and bound constraints.

### 2.6 Solution heuristic to the stochastic multimodal Stackelberg game

To solve $P_7$, we decompose $P_7$ into a lower-level problem, which is $P_6$, and an upper-level problem which solves for fares. The stochastic multimodal assignment game $P_6$ produces a set of matchings between users and operators ($f, y, v$) given a set of known fares ($\tilde{p}$). The fare adjusting problem solves for optimal fares ($p$) maximizing total revenue with a known set of path flows and operator decisions ($\tilde{f}, \tilde{y}, \tilde{v}$). We solve the 2 problems iteratively to solve $P_7$ as shown in **Figure 2**. Fares are initialized as all zeros. $P_6$ takes the solution of the fare adjusting problem ($p$) as the known fare parameters ($\tilde{p}$). The fare adjusting problem takes the solution of $P_6$ ($f, y, v$) as the known parameters to solve for a new set of fares ($p$). Details of the fare adjusting problem are introduced in section 2.6.1.

Another challenge is path set generation. To make sure that at convergence, the path set applied is reasonable, we do path set updates after solving $P_6$. A subnetwork identification process is integrated to eliminate unnecessary links and nodes to speed up the path set updates. Details are discussed in section 2.6.2.

We show that iterating the two problems leads to a stochastic Stackelberg equilibrium equivalent to $P_7$. Details are discussed in the rest of this section.

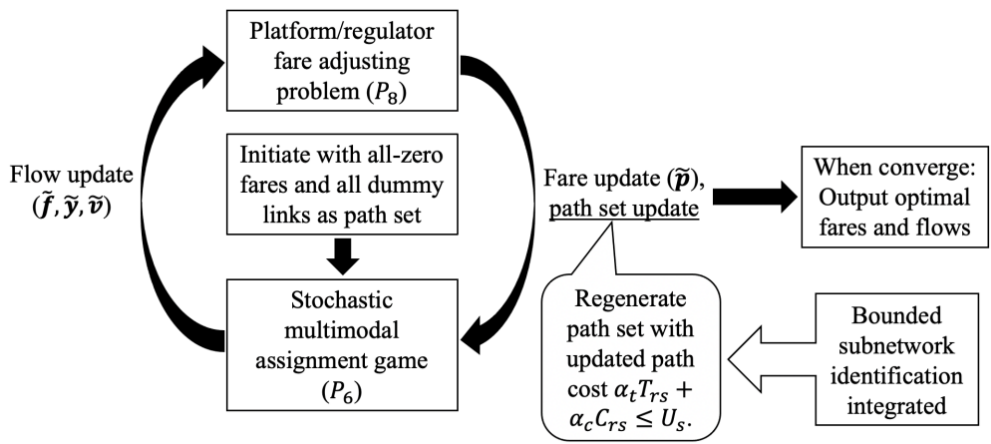

**Figure 2.** Solution heuristic to the stochastic multimodal Stackelberg game.

### 2.6.1  Fare adjustment

Although matchings of travelers and operators can be solved with fares known, fares cannot be solved with fixed path flows, since fare variables would be unbounded on the upper side. Operators try to gain as much as possible if path flows do not respond to fare changes. However, incorporating both path flows and fares as decision variables leads to a nonlinear objective function (Eq. (55a)) and nonlinear constraints (Eqs. (55b) – (55c)), which would make the fare adjustment problem hard to solve.

To tackle this issue, we first approximate total revenue with a solution from $P_6$. According to Eq. (38) and Eq. (40), path flows can be written as functions of fares Eq. (56), assuming that the Lagrange multipliers are fixed constants.

$$\tilde{f}_{rs}(\bm{p}) = \exp(-u_r - u_s) \tag{56}$$

Total revenue can be written as Eq. (57).

$$\mathcal{R}(\bm{p}) = \sum_{s \in S} \sum_{r \in R_s} \tilde{f}_{rs}(\bm{p}) \sum_{l \in A_r} p_l \tag{57}$$

$\mathcal{R}(\bm{p})$ can be interpretated as an approximation of the total revenue given pre-solved link/zone delays (as functions of Lagrange multipliers shown in **Proposition 2**) and travelers' payoffs $u_s$ solved from $P_6$. Maximizing $\mathcal{R}(\bm{p})$ with pre-solved Lagrange multipliers reflects how the regulator/platform would adjust the fares considering travelers' and operators' responses given a current system status, which is a solution from $P_6$. The anticipated responses are approximated with the delays and travelers' payoffs from the current system status.

Hence, to solve for the adjusted fares, we maximize $\mathcal{R}(\bm{p})$ with the following profitability constraints Eqs. (58) – (59). The two constraints are the same as constraints Eqs. (53b) – (53c), but with pre-solved $(\tilde{\bm{f}}, \tilde{\bm{y}}, \tilde{\bm{v}})$ plugged in.

$$\sum_{l \in A_f} \sum_{s \in S} \sum_{r \in R_s} \delta_{lrs} \tilde{f}_{rs} p_l \geq \sum_{l \in A_f} c_l \tilde{y}_l, \ \forall f \in Q_F \tag{58}$$

$$\sum_{l \in A_f} p_l \sum_{s \in S} \sum_{r \in R_s} \delta_{lrs} \tilde{f}_{rs} \geq \sum_{l \in A_f} \sum_{s \in S} m_l \sum_{r \in R_s} \delta_{lrs} \tilde{f}_{rs} + \sum_{i \in N_f} q_i \tilde{v}_i, \ \forall f \in Q_M \tag{59}$$

To make the fare adjustment problem bounded, an upper bound constraint Eq. (60) is added. Constraint Eq. (60) makes sure that traveler's path costs are upper-bounded. Link and node delays can be excluded, since the functionality of the constraint is just to provide an upper bound of the fares. $\mathcal{B}_{rs}$ is a pre-defined parameter, which is the upper bound. $\mathcal{B}_{rs}$ can be defined as a reasonably large number, which is larger than $U_s$ to give more space for fare adjustment of a single iteration, $1.5 U_s$ or $2 U_s$ are reasonable values.

$$\sum_{l \in A_r \cap (A_O \cup A_F \cup A_D \cup A_M \cup A_{OM})} t_l + \sum_{l \in A_r} p_l \leq \mathcal{B}_{rs}, \ \forall r \in R_s, s \in S \tag{60}$$

The fare adjustment problem $P_8$ is formulated as follows.

$$P_8 : \max_{p} \mathcal{R}(\bm{p})$$

subject to Eq.(56) – (58) and (53f).

$P_8$ has a nonlinear objective function. We apply the Frank-Wolfe algorithm (Frank and Wolfe, 1956) to adjust fares with known path flows and operator decisions $(\tilde{f}, \tilde{y}, \tilde{v})$ solved from $P_6$. First, we find the gradient at the point of the current solution of $P_6$ by taking the derivative of $\mathcal{R}(p)$ w.r.t. $p_l$ as shown by Eq. (61), where $R^{ls}$ denotes the set of paths of traveler group $s \in S$ that incorporates link $l$.

$$\frac{\partial \mathcal{R}(p)}{\partial p_l} = \sum_{s \in S} \sum_{r \in R^l} \tilde{f}_{rs}(-(\alpha_t - \alpha_c)p_l + 1), \forall l \in A_r, r \in R_s, s \in S \tag{61}$$

Plugging in the fares $\tilde{p}_l$ that are used to solve $P_6$, we have a vector of gradient $\nabla \mathcal{R} = [\frac{\partial \mathcal{R}(p)}{\partial p_l}, \forall l \in A_r, r \in R_s, s \in S]$. Then we solve the following linear subproblem $P_9$ to find the intersection point $p'_l$ where the steepest direction hits the boundary of the feasible region.

$$P_9: \max_{p'} \sum_{l \in A_r, r \in R_s, s \in S} \frac{\partial \mathcal{R}(p)}{\partial p_l} p'_l \tag{62}$$

subject to Eqs. (55) – (57) and (52f)

but with all decision variables $p_l$ replaced by $p'_l$.

Then we do a line search on the line between the current fares $\tilde{p}_l$ and $p'_l$ to find the point on the lines which maximize $\mathcal{R}(p)$. We do that through solving $\beta^* = argmax\, \mathcal{R}(\tilde{p} + \beta(p' - \tilde{p}))$, where $0 \leq \beta \leq 1$. Then we update fares with the optimal $\beta$: $\tilde{p} = \tilde{p} + \beta^*(p' - \tilde{p})$.

The traditional Frank-Wolfe algorithm iterates the above process to obtain an optimal solution of a nonlinear program. However, in this case, the fare update is solved assuming suboptimal matching solution $(\tilde{f}, \tilde{y}, \tilde{v})$ and corresponding Lagrange multipliers $u, \gamma, \pi$. Hence, it is not worthwhile to iterate to convergence to update fares every time after solving $P_6$. Instead, it can be iterated for a preset $n$ times as maximum. In the numerical examples shown in section 3, we only do 1 iteration of Frank-Wolfe algorithm for each fare adjustment, and it is still shown to be effective. The algorithm of fare adjustment is summarized as **Algorithm 2**. Convergence tolerance is denoted as $\varepsilon$.

**Algorithm 2.** Frank-Wolfe algorithm for fare adjustment.

---
**Inputs:** current fares $\tilde{p}$; solution of $P_6$: $(f, y, v)$; Lagrange multipliers of $P_6$: $u, \gamma, \pi$; path set $R_s$, $s \in S$, network parameters: $t, m, c, w, q, z$,; demand: $d$; weights: $\alpha_t, \alpha_c$.
$itr = 0$.
$(\tilde{f}, \tilde{y}, \tilde{v}) = (f, y, v)$.
**While** $itr \leq n$:
    **Step 1 (Gradient):** Find gradient $\nabla \mathcal{R} = [\frac{\partial \mathcal{R}(p)}{\partial p_l}, \forall l \in A_r, r \in R_s, s \in S]$ with Eq. (61).
    **Step 2 (Linear subproblem):** Solve $P_9$ with the gradient $\nabla \mathcal{R}$ from Step 1.
    **Step 3 (Line search):** Solve $\beta^* = argmax\, \mathcal{R}(\tilde{p} + \beta(p' - \tilde{p}))$, where $0 \leq \beta \leq 1$.
    **Step 4 (Fare and iteration number update):** $\tilde{p} = \tilde{p} + \beta^*(p' - \tilde{p})$, $itr = itr + 1$.
---

**Step 5** (Convergence check): **if** $\beta^* < \varepsilon$, **break**.
**Outputs:** $\tilde{p}, \beta^*$.

*2.6.2 Path set update with subnetwork identification*

Considering all possible matchings means path enumeration, which would not be computationally effective for real-sized networks. According to Eq. (36), path flows solved from $P_6$ are distributed according to path disutilities $u_{rs} = -\alpha_t T_{rs} - \alpha_c C_{rs}, \forall r \in R_s, s \in S$. The path set should also be identified according to $u_{rs}$. To avoid path enumeration, we define $\alpha_t U_s$ as the upper bound of $-u_{rs}$ of all $R_s, \forall s \in S$. Adopting such a bound brings back the assumption of deterministic assignment games mentioned in section 2: a matching will not form if there is no positive gain to be shared by the seller (operators) and buyer (travelers). As shown by Eq. (22), the shared gain $a_{rs}$ of a trip on path $r \in R_s, s \in S$, can be negative, zero, or positive. However, after adopting the path cost bound, $a_{rs}$ becomes strictly positive. Characteristic function $a_{rs}$ with path cost bound $\alpha_t U_s$ is equivalent to Eq.(63), where $\bar{c}_l$ is operating cost of link $l$ per traveler. It is obvious that the computation is equivalent to the characteristic function computation of deterministic assignment games.

$$a_{rs} = \max\left(0, \alpha_t U_s - \sum_{l \in A} \delta_{lrs}[\alpha_t(t_l + p_l) + \alpha_c(\bar{c}_l - p_l)]\right), \quad \forall r \in R_s, s \in S \quad (63)$$

We aim to have path sets $R_s^*$ of traveler group $s \in S$ which includes exactly all the paths with trip utility $-u_{rs} \leq \alpha_t U_s$ at convergence. Since path disutilities $u_{rs}$ change with the updates of fares and flows, the path set needs to be updated with fares and flows. We do a path set generation after solving $P_6$ with a ***bounded k-shortest path finding algorithm*** to find $R_s$ for all the OD pairs of traveler group $s \in S$. The bounded k-shortest path finding algorithm is a slightly modified version of the k-shortest path finding algorithm proposed by Yen (1971). We added a simple cycle check and a bound check. The algorithm is done separately for each OD pair. Paths with repeated nodes are deleted after found. The algorithm ends for an OD pair when a path with $-u_{rs} \geq \alpha_t U_s$ is found. If the final path has $-u_{rs} > \alpha_t U_s$, it is deleted from the path set.

However, if the path set generation is done in every iteration after $P_6$ is solved, it becomes very computationally expensive with real-sized networks. Instead of doing it every iteration, we use the following criteria to determine whether a path set generation is needed:

- When link flow change or link fare change (L2 norm of the change vector) is larger than a criterion $\theta$, a path set generation is done since the disutility change is significant enough;
- When link flow change or link fare change (L2 norm of the change vector) is smaller than the convergence criterion $\epsilon$, a path set generation is done to check if the path set converges.

The above path set update strategy is integrated into the overall solution heuristic in section 3.6.3 (**Algorithm 4**).

To further reduce computation cost of the bounded k-shortest path finding algorithm, we apply a subnetwork identification approach, as presented in the **Algorithm 3**.

**Algorithm 3.** Subnetwork identification.

---

**Inputs:** Origin node: $O$; destination node: $D$; network graph $G$; network parameters: $\boldsymbol{t, m, c, w, q, z}$; current fares $\boldsymbol{p}$; solution of $P_6$: $(\boldsymbol{f, y, v})$; Lagrange multipliers of $P_6$: $\boldsymbol{u, \gamma, \pi}$.
$G' = G$.
Compute link costs $-u_l, \forall l \in A_r, r \in R_s, s \in S$ with $\boldsymbol{f, y, v, p, \gamma, \pi}$.
Do Dijkstra's algorithm from $O$ to all nodes $i$ in $G$, shortest distance from $O$ to $i$ is $\tau_{Oi}$.
Do Dijkstra's algorithm from $D$ to all nodes $i$ in $G$ with all link directions reversed, shortest distance from $D$ to $i$ is $\tau_{Di}$.
**For** node $i$ in $G$:   #node elimination
  **If** $\tau_{Oi} + \tau_{Di} > \alpha_t U_s$, delete node $i$ and all links connected with node $i$ from $G'$.
**For** link $(i, j)$ in $G$:   #link elimination
  **If** $\tau_{Oi} + u_{(i,j)} + \tau_{Dj} > \alpha_t U_s$, delete link $(i, j)$ from $G'$ if link $(i, j)$ is in $G'$.
**Output:** $G'$.

---

What remains in $G'$ is a minimal subnetwork containing all paths that satisfy the bound, while not all paths in the subnetwork will satisfy the bound. **Proposition 4** establishes the optimality of **Algorithm 3** in obtaining a minimal subnetwork.

**Proposition 4.** Minimal subnetwork. *Algorithm 3 outputs a minimal subnetwork which contains all the paths that satisfies the bound $\alpha_t U_s$.*

***Proof.*** For a node $i$ in $G$, $\tau_{Oi} + \tau_{Di} > \alpha_t U_s$ means that the shortest path from $O$ to $D$ that traverses $i$ is above the bound $\alpha_t U_s$. This means that all the paths from $O$ to $D$ that traverse $i$ are above the bound. Hence node $i$ would not be on any paths from $O$ to $D$ below the bound, so node $i$ and all links connected to it can be removed from $G$ without impacting the path generation. $\tau_{Oi} + \tau_{Di} \leq \alpha_t U_s$ means that there exists one or more paths $r$ in $G$ that traverse $i$ with $-u_{rs} \leq \alpha_t U_s$, hence node $i$ is a part of the minimal subnetwork that contains all the paths $r$ with $-u_{rs} \leq \alpha_t U_s$.

For a link $(i, j)$ in $G$, $\tau_{Oi} + u_{(i,j)} + \tau_{Dj} > \alpha_t U_s$ means that the shortest path from $O$ to $D$ that traverse $(i, j)$ is above the bound $\alpha_t U_s$. Similarly, link $(i, j)$ can be removed from $G$ without impacting the path generation. $\tau_{Oi} + u_{(i,j)} + \tau_{Dj} \leq \alpha_t U_s$ means that there exists one or more paths $r$ in $G$ that traverse $(i, j)$ with $-u_{rs} \leq \alpha_t U_s$, hence link $(i, j)$ is a part of the minimal subnetwork that contains all the paths $r$ with $-u_{rs} \leq \alpha_t U_s$. ∎

### 2.6.3   Solution heuristic to the stochastic multimodal Stackelberg game

The following **Algorithm 4** is a proposed heuristic to solve the stochastic multimodal Stackelberg game. We denote link flows by $\boldsymbol{x}$. **Proposition 4** shows that the output of **Algorithm 4.4** is equivalent to the optimum of $P_7$ when it converges. As shown in **Figure 4.2**, we do iterative updates of the fares and stochastic multimodal assignment game, with the path set updated as described in section 2.6.2.

**Algorithm 4.** Solution heuristic to the stochastic multimodal Stackelberg game.

---

**Inputs:** network parameters: $\boldsymbol{t, m, c, w, q, z}$; demand: $\boldsymbol{d}$; weights: $\alpha_t, \alpha_c$.
**Step 1.** (Initialize):

        $\tilde{p} = 0$ (link fares).
        $R_s = [[\ell_s]], s \in S$ (path sets: initialized with dummy links).
        $f^{prev} = 0, y^{prev} = 0, v^{prev} = 0, p^{prev} = 0, x^{prev} = 0$ (track previous iterations).
**Step 2.** (Iterate):
    Iterate until path flows and path sets converge:
        **Step 2.1.** (Assignment game):
            Solve $P_6$ with **Algorithm 1** with $\tilde{p}$ and $R_s$ to obtain $(f, y, v)$ and $x, \gamma, \pi$.
            $(\tilde{f}, \tilde{y}, \tilde{v}) = (f, y, v)$.
        **Step 2.2.** (Fare adjustment):
            Solve $P_8$ with **Algorithm 2** with $(\tilde{f}, \tilde{y}, \tilde{v})$ and $R_s$ to obtain $p$.
            $\tilde{p} = p$.
        **Step 2.3.** (Path set update):
            Update link disutilities with $f, y, v, p, \gamma, \pi$.
            If $||x^{prev} - x|| < \epsilon$ or $||x^{prev} - x|| > \theta$ (or compare $p^{prev}$ and $p$):
                For $s \in S$:
                    Apply **Algorithm 3** with $\tilde{f}, \tilde{y}, \tilde{v}, \tilde{p}$ to obtain subnetwork $G'$.
                    Do a bounded k-shortest path finding with bound $\alpha_t U_s$ and $G'$ to find $R_s$.
            If $||x^{prev} - x|| < \epsilon$ and path set $R_s, \forall s \in S$ does not change:
                **Break**   #convergence
        **Step 2.4.** (Update tracking variables):
            $(f^{prev}, y^{prev}, v^{prev}) = (f, y, v), p^{prev} = p, x^{prev} = x$.
**Outputs:** $f; y; v; p; R_s, \forall s \in S$.

**Proposition 4.** Bilevel equivalence. *Algorithm 4 obtains a Stackelberg equilibrium which is equivalent to the solution to $P_7$.*

***Proof.*** When Step 2 of **Algorithm 4** converges, $\tilde{f}, \tilde{y}, \tilde{v}, \tilde{p}$, Lagrange multipliers $\gamma, \pi$, and path set $R_s, \forall s \in S$ all stay stable over iterations. Denote the stable values as $\bar{f}, \bar{y}, \bar{v}, \bar{p}, \bar{\gamma}, \bar{\pi}$. They satisfy optimality conditions of both $P_6$ and $P_8$. $P_6$ is equivalent to 4 constraints of $P_7$, which are Eqs. (53d, 53e, 53g, 53h). $P_8$ approximates the rest of $P_7$ with assumed Lagrange multipliers $\gamma, \pi$. When $\bar{f}, \bar{y}, \bar{v}, \bar{p}, \bar{\gamma}, \bar{\pi}$ satisfy the optimality conditions of $P_6$, constraints Eqs.(53d, 53e, 53g, 53h) are satisfied. Since $\bar{f}, \bar{y}, \bar{v}, \bar{\gamma}, \bar{\pi}$ stay stable over iterations, $P_8$ is no longer an approximation but solves for optimal fares $\bar{p}$. Hence when Step 2 of **Algorithm 4** converges, $P_6$ and $P_8$ iteratively gives the optimal solution of $P_7$. ∎

    As mentioned in section 2.6.1, **Algorithm 2** does not need to be fully converged in each iteration of **Algorithm 3**. Instead, a maximum number of iterations $n$ can be set. When **Algorithm 3** approaches convergence, the changes of $p$ would be very small, leading to immediate convergence of **Algorithm 2**. In application, a special case is $n = 1$, which means only one iteration is done in each run of **Algorithm 2**. In this case, $\beta^*$ found in the only iteration indicates the change of fares in the iteration of **Algorithm 3**, hence can be used to check convergence in Step 3 in **Algorithm 3** to replace $||x^{prev} - x||$ or $||p^{prev} - p||$. This convergence check is applied in the numerical tests in section 3.

# 3 Numerical experiments

## 3.1 Illustrative Example

We use a small network shown in **Figure 3** to first illustrate the model. There are two OD pairs: 1 to 3 (traveler group 1) and 1 to 2 (traveler group 2), both with 100 units of travel demand. There is a bus operator considering providing a service connecting 1 and 2, with a travel cost $t_{12,B} = \$4$, an operation cost $c_{12,B} = \$300$, and a maximum capacity $w_{12,B} = 50$. There are 2 unowned access (walking) links connecting 1 and 3 (time cost $t_{13,w} = \$20$), and 2 and 3 (time cost $t_{23,w} = \$6$). To represent travelers who choose not to use the platform, a dummy link is added for each OD pair. The cost of this link corresponds to the pre-estimated disutility of the best external option, set to $U_s = \$15$ for both traveler groups, reflecting the perceived generalized cost of alternative modes such as private car, competing platforms, or not traveling. There is also a MoD operator considering operating within a region composed of 3 zones: A,B,C. The centroids of zone A,B,C are assumed to be nodes 1,2,3, respectively. Travelers can make transfers between bus/walking and MoD at nodes 1,2,3 if MoD services are provided in the corresponding zone. MoD travel cost between zone A and B, B and C, and A and C are \$8, \$3, and \$10, respectively. The original network shown in **Figure 3(a)** is expanded into the network in **Figure 3(b)**. MoD nodes which correspond to centroids 1,2,3 are A,B,C, respectively. MoD nodes are connected with each other as a complete network (blue links), and are connected with the corresponding centroids with bidirectional MoD access/egress links (green links). Uncongested access cost of all MoD nodes are \$5. Travel costs of MoD egress links are 0 for all MoD egress links. Operation cost of the MoD operator include the cost of operating the links connecting A,B,C, and the infrastructure cost at A,B,C. Operation cost of the MoD links per unit demand is $m_l = \$0.2$. MoD fleet deployment cost of zone A,B,C are $q_i = \$20$. Maximum fleet size for zone A,B,C are all $z_i = 200$. There are 6 nodes and 17 links (include 2 dummy links) in the expanded network.

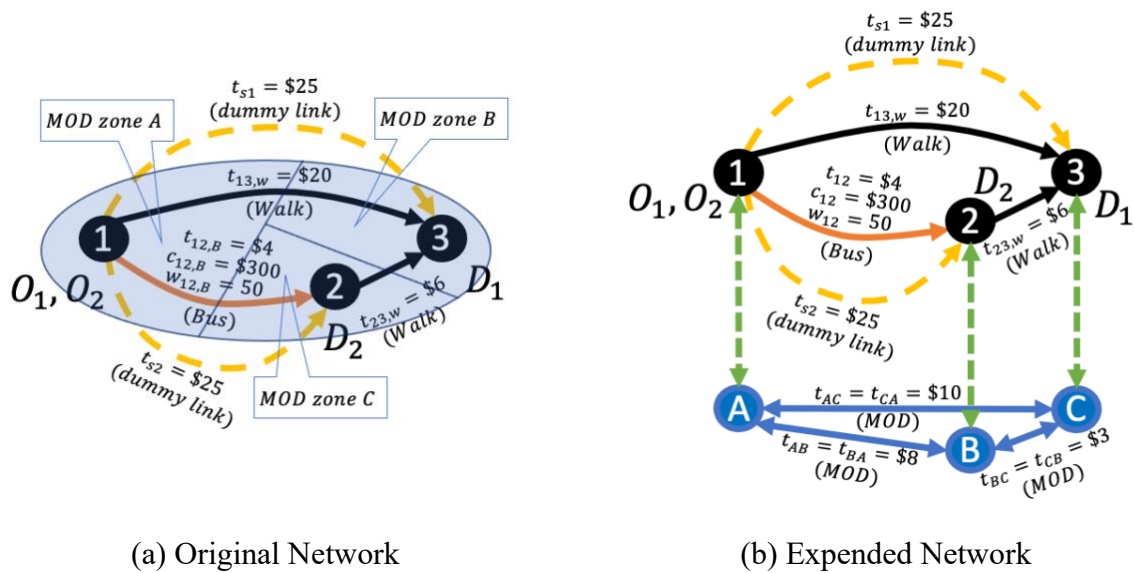

(a) Original Network  (b) Expended Network

**Figure 3.** Illustrative Example Network.

This illustrative example is solved with Python 3.7 on a laptop with 2.3 GHz Quad-Core Intel Core i7 and 32 GB 3733 MHz LPDDR4X memory with **Algorithm 4**, converged after 5 iterations. Weights are set as $\alpha_t = 1$ and $\alpha_c = 0.5$, indicating that operator costs are weighted at half the level of traveler costs in this illustrative example. **Algorithm 1** converges when the change of L2 norm of flow vector is smaller than $10^{-4}$. **Algorithm 2** runs for 1 iteration every time. When $\beta^* > 0.05$, fare changes are large enough to trigger a path set update. When $\beta^* < 10^{-4}$, the final path set check is triggered: when the path set does not change, **Algorithm 4** converges.

**Table 5** shows the total revenue, $\beta^*$, number of paths found, and number of links and nodes removed by **Algorithm 3**. Total revenue converged to $358.70 with a CPU time of 2.63 sec. Total travelers' expected payoff is 1490.28, among which OD 1 is $0 and OD 2 is 1490.28. At convergence, 3 paths are used for OD 1; 2 are used for OD 2. Path set generation is performed in iteration 0,1,3,4, in which iteration 0,1,3 are triggered by $\beta^* > 0.05$, iteration 4 is triggered by $\beta^* < 10^{-4}$. The whole network is removed in iteration 1,3,4 for OD 1, hence only dummy link is used by travelers of OD 1 in those iterations.

**Table 6** shows the output of **Algorithm 4**. All the travelers of both OD 1 are on dummy link, meaning that they are not making the trip or not using the platform. Transit link (1,2) provides the maximum capacity. The capacity constraint of transit link (1,2) binds, leading to a delay of $4.29 per traveler. For MoD, zone B and C are chosen to be operated, serving travelers of OD 2. None of them provides the maximum fleet size and none of them has delay. Disutilities of the used paths are shown in **Table 6**, which could verify the logit path flows.

Table 5. Summary of the iterations of **Algorithm 4**.

| Iteration | Revenue ($) | $\beta^*$ | Number of paths (exclude dummy links) | | Subnetwork Identification | | | |
|---|---|---|---|---|---|---|---|---|
| | | | | | Number of nodes removed | | Number of links removed | |
| | | | OD 1 | OD 2 | OD 1 | OD 2 | OD 1 | OD 2 |
| 0 | 0 | 1.000 | 1 | 2 | 3 | 2 | 12 | 8 |
| 1 | 376.60 | 0.073 | 0 | 2 | 6 | 2 | 15 | 8 |
| 2 | 356.58 | 0.013 | - | 2 | - | - | - | - |
| 3 | 350.83 | 0.151 | 0 | 2 | 6 | 2 | 15 | 8 |
| 4 | 358.70 | 0.000 | 0 | 2 | 6 | 2 | 15 | 8 |

Table 6. Results of the small example.

| OD pair s | Used paths | | | | | Link fares | |
|---|---|---|---|---|---|---|---|
| | Path r | $f^*_{rs}$ | $-u_{rs}$ | $u_s$ | | Link l | $p^*_l$ ($) |
| OD 1 | 1→3 (dummy) | 100 | 15 | 19.61 | | (1,2) | 4.34 |
| OD 2 | 1→2 (dummy) | 16.21 | 15 | | | (A,B) | 0 |
| | 1→2 | 50 | 13.87 | 17.79 | | (B,A) | 0 |
| | 1→A→B→2 | 33.79 | 14.27 | | | (B,C) | 4.19 |
| **Fixed-route links** | | | **MoD zones** | | | (C,B) | 0 |
| Link l | $y^*_l w_l$ | $D^{A_F*}_l$ ($) | Zone i | $v^*_i z_i$ | $D^{N_M*}_l$ ($) | (A,C) | 0 |
| (1,2) | 50 (max capacity) | 4.29 | A | 0 | 0 | (C,A) | 0 |
| | | | B | 33.79 | 0 | | |
| | | | C | 33.79 | 0 | | |

To verify that the path set found at convergence are all the paths below the bound $\alpha_t U_s = \$15$, we enumerate all the unused paths and their costs at convergence in **Table 7**. None of them has a cost below the bound.

Table 7. Unused path disutilities.

| OD pair $s$ | Unused Path $r$ | $-u_{rs}$ |
|---|---|---|
| OD 1 | 1→2→B→C→3 | 22.09 |
| | 1→A→C→3 | 15.23 |
| | 1→2→3 | 19.87 |
| | 1→A→B→C→3 | 17.42 |
| | 1→A→B→2→3 | 20.27 |
| OD 2 | 1→3→C→B→2 | 28.22 |

*3.2 Sioux Falls Example*

The model is further tested on an expanded Sioux Falls network. This Sioux Falls network is considered as a combination of a walking/transfer network and fixed-route transit service links. Network parameters (link costs and capacities) are shown in **Appendix A**. Walking links have 0 operation cost and are not owned by any operator. All fixed-route transit links have an operation cost of $200.

There are 4 fixed transit lines in this case, marked in **Figure 4** with blue (Operator 1), pink (Operator 2), yellow (Operator 3), and green (Operator 4). All other links are walking/transfer links. The four fixed-route operators can choose a capacity to provide below or equal to their maximum capacities (can be 0) on each of the links they own.

Three MoD operators operate in the region, marked with purple (Operator 5), light blue (Operator 6), and orange (Operator 7). Candidate service regions that they cover are marked in **Figure 4**. The maximum fleet size to be deployed in each MoD zone is 5,000. The cost of deploying the maximum fleet size in each MoD zones is $200.

Travel demand is shown in **Appendix B**. The OD pairs are all the combinations between nodes 1, 2, 12, 18, 13, and 20 (30 OD pairs). Travelers of each OD pair is a traveler group. Total demand is 9,700. Pre-estimated disutility of the best external choice $U_s$ is $30 for all OD pairs. The access/wait cost function and operation cost function of MoD are the same as the illustrative case. The uncongested access/wait cost of MoD is $1, while the operation cost of the MoD links per unit demand is $1.

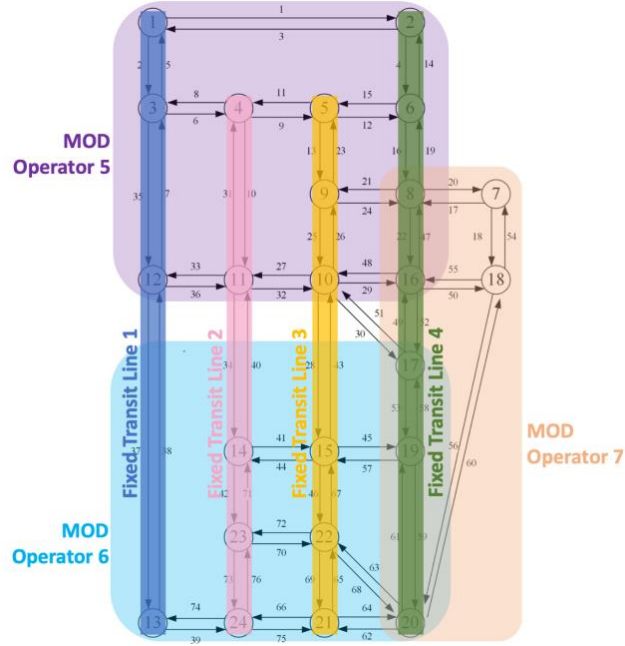

**Figure** Error! No text of specified style in document.. Network construction of the Sioux Falls case.

The expanded network has 53 nodes and 426 links. All the nodes covered by each MoD operator's service regions are mutually connected with MoD links. The travel costs of MoD links are set to be 75% of the cost of the shortest path between the 2 nodes on the original network. **Algorithm 1** converges when the change of L2 norm of flow vector is smaller than $10^{-4}$. **Algorithm 2** runs for 1 iteration every time. When $\beta^* > 0.05$, fare changes are large enough to trigger a path set update. When $\beta^* < 10^{-3}$, the final path set check is triggered: when path set does not change, **Algorithm 4.4** converges.

Two cases are run to test the impact of $\alpha_t$ and $\alpha_c$. For Case 1, $\alpha_t = 1$ and $\alpha_c = 0.5$. For Case 2, $\alpha_t = 1$ and $\alpha_c = 0.75$. Case 2 has higher weights for operators' valuations than Case 1, which should lead to higher fares and operator revenue at equilibrium. All cases are run on the same machine as the small case.

### 3.2.1 Case 1

The CPU time of Case 1 ($\alpha_t = 1, \alpha_c = 0.5$) is 5h 38min. **Algorithm 4** converged after 61 iterations, in which 20 iterations did path set generation. Due to the complete graphs, the results are not plotted but listed in tables: link flows, fares, and capacities of fixed-route links are shown in **Appendix C**; link flows and fares of MoD links are shown in **Appendix D**; link flows of walking links are shown in **Appendix E**; capacities of MoD zones are shown in **Appendix F**. Summary of the results is in **Table 5**. Revenue of convergence is $15127.13. Travelers' total expected payoff is $151076.68. Unserved demand (total flow on dummy links) is 624.23. None of the fixed-route links has binding maximum capacity. Neither do the MoD zones. All capacities/fleet sizes provided equal to the amount of flow traversing the fixed-route link/MoD node. There is no delay in the network. The trajectories of total revenue and $\beta^*$ are shown in **Figure 5** and **6**.

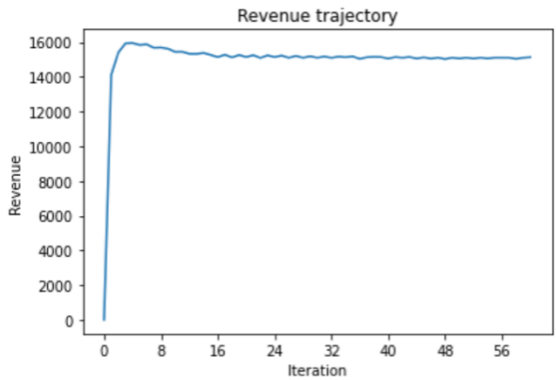

**Figure 5.** Revenue trajectory of Case 1 ($\alpha_t = 1$, $\alpha_c = 0.5$).

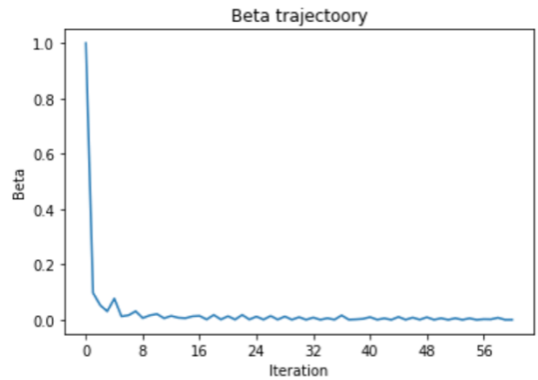

**Figure 6.** $\beta^*$ trajectory of Case 1 ($\alpha_t = 1$, $\alpha_c = 0.5$).

The percentage of nodes and links remained after subnetwork identification (**Algorithm 3**) in each iteration with path generation is shown in **Figure 7** and **8**. For most OD pairs, more than 50% of the network is removed. To show the impact of subnetwork identification on the efficiency of path set generation, we run Case 1 without subnetwork identification and compare the run time of path set generation of each iteration. The comparison is shown in **Figure 9**. Path generation is significantly reduced due to network size reduction, which shows the effectiveness of subnetwork identification.

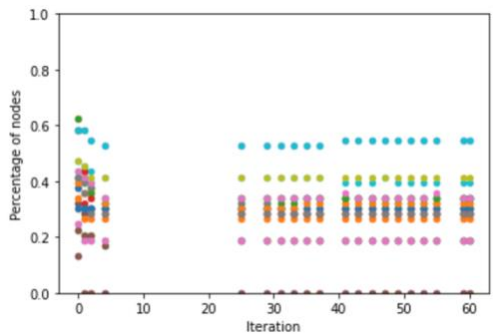

**Figure 7.** Percentage of nodes remained after subnetwork identification (colors represent OD pairs) (Case 1: $\alpha_t = 1$, $\alpha_c = 0.5$).

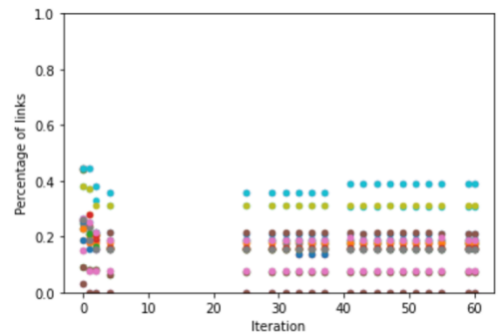

**Figure 8.** Percentage of links remained after subnetwork identification (colors represent OD pairs) (Case 1: $\alpha_t = 1$, $\alpha_c = 0.5$).

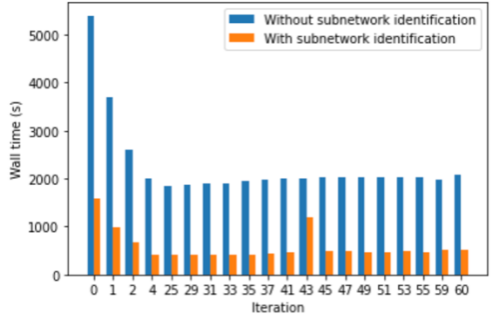

**Figure 9.** Path set generation time comparison: with vs. without subnetwork identification (Case 1: $\alpha_t = 1$, $\alpha_c = 0.5$).

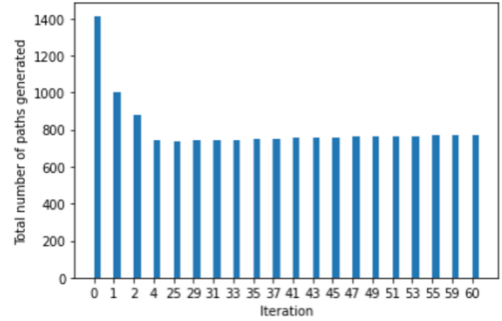

**Figure 10.** Total number of paths generated each iteration (Case 1: $\alpha_t = 1$, $\alpha_c = 0.5$).

*3.2.2 Case 2*

CPU time of Case 2 ($\alpha_t = 1$, $\alpha_c = 0.75$) is 1h 16min. **Algorithm 4** converged after 18 iterations, in which 6 iterations performed path set generation. Summary of the results is in **Table 5**. Revenue of convergence is $29717.73. Travelers' total expected payoff is $156308.70. Unserved demand (total flow on dummy links) is 653.46. Same as Case 1, none of the fixed-route links and MoD zones have binding maximum capacity. There is no delay in the network. The trajectories of total revenue and $\beta^*$ are shown in **Figure 11** and **12**.

**Table 8.** Summary of Case 1 and Case 2.

|  | Operator | Number of links/zones operated |  | Flow |  | Revenue ($) |  | Travelers' total expected payoff ($) |  |
|---|---|---|---|---|---|---|---|---|---|
|  |  | Case 1 | Case 2 | Case 1 | Case 2 | Case 1 | Case 2 | Case 1 | Case 2 |
| **Fixed-route** | 1 (Blue line) | 6 | 6 | 6728.64 | 6735.57 | 8021.61 | 16578.93 | 138716.9 | 140132.4 |
|  | 2 (Pink line) | 12 | 12 | 488.05 | 481.65 | 135.29 | 319.66 |  |  |
|  | 3 (Yellow line) | 2 | 2 | 98.12 | 75.09 | 97.93 | 167.76 |  |  |
|  | 4 (Green line) | 4 | 4 | 260 | 212.18 | 18 | 325.92 |  |  |
|  | Total | 24 | 24 (+0%) | 7574.81 | 7504.49 (-0.9%) | 8272.82 | 17392.27 (+110.2%) |  |  |
| **MoD** | 5 (Purple MoD) | 62 | 58 | 2596.51 | 2484.83 | 3013.74 | 5170.03 |  |  |
|  | 6 (Blue MoD) | 58 | 50 | 2102.93 | 1929.63 | 2601.15 | 5045.95 |  |  |
|  | 7 (Orange MoD) | 36 | 32 | 856.66 | 758.04 | 1081.54 | 2092.49 |  |  |
|  | Total | 156 | 140 (-10.3%) | 5556.1 | 5172.5 (-6.9%) | 6696.43 | 12308.47 (+83.8%) |  |  |
| **Walk** |  | - | - | - | 4850.01 | 5109.21 (+5.3%) | - | - |  |

Compared with Case 1, Case 2 has higher weights for operators' valuations, leading to significantly higher optimal revenue (96.5% increase) while the unserved demand does not drop much (4.7% increase). As shown in **Table 8**, the total flows of both fixed-route services and MoD services both drop, service coverage of MoD becomes smaller, while revenue increase significantly. The reason is that travelers' costs become less important in the coalitional logit model and demand is less sensitive to fares. Case 2 converges much faster than Case 1, since Case 2 has a significantly smaller path set, leading to less iterations of path set check when the convergence criterion ($\beta^* < 10^{-3}$) is met. Note that travelers' total expected payoffs are not comparable with different $\alpha_t, \alpha_c$.

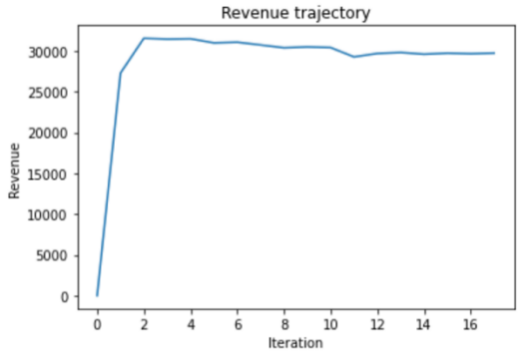 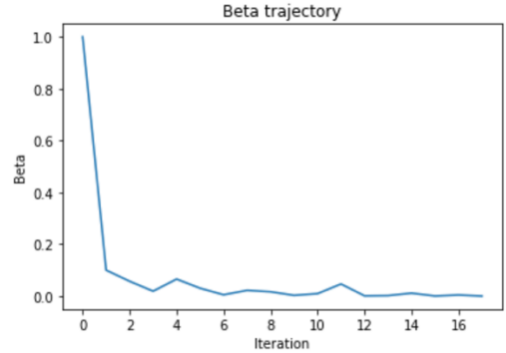

**Figure 11.** Revenue trajectory of Case 2 ($\alpha_t = 1, \alpha_c = 0.75$).

**Figure 12.** $\beta^*$ trajectory of Case 2 ($\alpha_t = 1, \alpha_c = 0.75$).

The percentage of nodes and links remaining after subnetwork identification in each iteration with path generation is shown in **Figure 13** and **14**. Similar to Case 1, more than 50% of the network is removed for most OD pairs. **Figure 15** shows the path set generation time of each iteration. **Figure 16** shows the total number of paths generated for all OD pairs in each iteration. Compared with Case 1, the path set generation time is reduced. The reason is that the increase in $\alpha_c$ leads to higher $u_{rs}$ overall, more links and nodes are removed through subnetwork identification. The number of paths in the path set is also smaller.

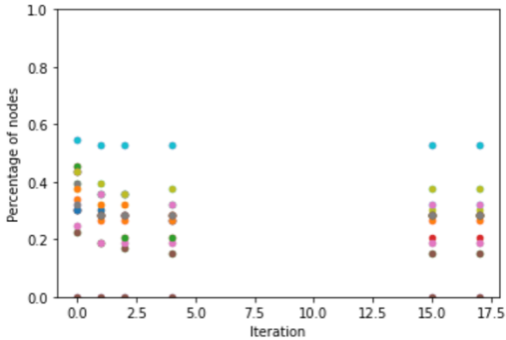 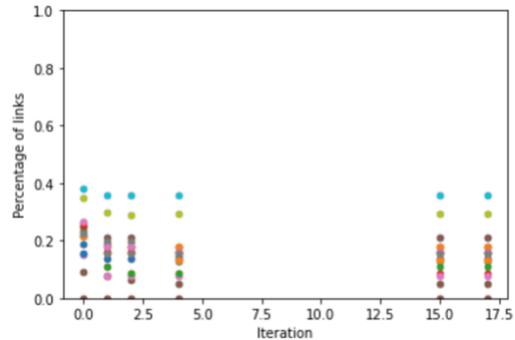

**Figure 13.** Percentage of nodes remained after subnetwork identification (colors represent OD pairs) (Case 1:$\alpha_t = 1, \alpha_c = 0.75$).

**Figure 14.** Percentage of links remained after subnetwork identification (colors represent OD pairs) (Case 1:$\alpha_t = 1, \alpha_c = 0.75$).

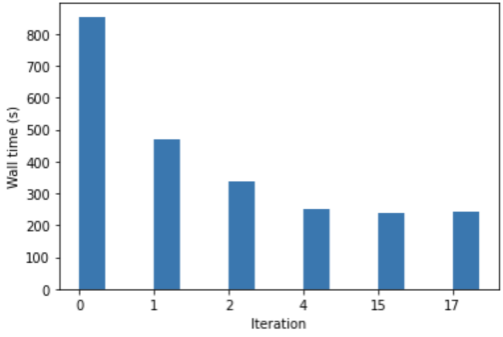
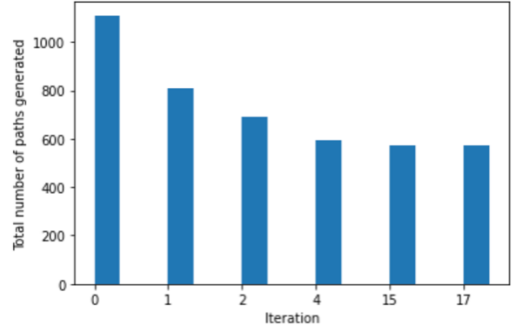

**Figure 15.** Path set generation time (Case 2: $\alpha_t = 1$, $\alpha_c = 0.75$).

**Figure 16.** Total number of paths generated each iteration (Case 2: $\alpha_t = 1$, $\alpha_c = 0.75$).

## 4 Conclusion

We study stochastic assignment games and its extensions in urban mobility markets. General forms of one-to-one and many-to-many stochastic assignment games are presented. Optimality conditions are discussed. The core of stochastic assignment games is defined in **Theorem 1**. To apply stochastic assignment games to the urban mobility markets, we extend the general stochastic many-to-many assignment game into a stochastic multimodal assignment game. Based on that, we propose a stochastic multimodal Stackelberg game as a bilevel problem to model urban mobility markets, in which the regulator/MaaS platform is the leader, travelers and operators are the followers, considering fixed-route and MoD operators. The regulator/platform maximizes revenue through fare setting from the upper level. Travelers and operators react to the fares, forming stochastic many-to-many matchings with stochasticity on the lower level.

An iterative balancing algorithm is proposed to solve the stochastic multimodal assignment game (lower-level problem). The bilevel problem is solved through an iterative fare adjusting heuristic, whose solution is shown to be equivalent to the bilevel problem with on condition when it converges. In each iteration, fare adjustment is done through a Frank-Wolfe algorithm. To avoid path enumeration, path set update is performed in an iterative way to have a bounded path set at convergence, with subnetwork identification to speed it up. Two case studies are conducted.

The study has several major contributions. First, the classic assignment game is extended to a stochastic form in which buyers and sellers have mixed strategies for choosing matches to form, where the probability is linked to utility functions. Second, the many-to-many assignment game of MaaS is extended into a similar stochastic form and solved. This form includes stochastic decisions for traveler path choices as well as operator route service choices. An exact solution method is provided based on iterative matching and proven to converge. Third, the role of the platform is explored through the Stackelberg game solved with a heuristic method to tackle the challenges of bilevel optimization with bounded path sets. The model could be applied to model, evaluate, and design MaaS systems from the perspective of the platform. Public agencies could also make use of the model to help evaluate and manage multimodal transportation systems.

One possible next step of the work would be solving the Stackelberg game with alternative objectives. The objective adopted in the proposed stochastic multimodal Stackelberg game. However, regulators might have various objectives when designing fares, such as maximizing social welfare (total traveler payoff), sustainability, resilience, and equity. Another possible direction is to study how different levels of heterogeneity affect the stability of the resulting allocations, especially when perception differences or operator cost variations become large. Other solution algorithms might need to be considered for the upper-level problem to account for various objectives. Link-based formulations, different types of subscription plans for pricing schemes, and other control variables beyond pricing (hub locations) will also be explored.

**Acknowledgments**

The authors were partially supported by the C2SMARTER Center, USDOT #69A3551747124. We thank Martin Strehler[1] for suggesting the subnetwork identification idea used in Algorithm 3.

---

[1] Professor of Mathematics, Department of Mathematics, Westsächsische Hochschule Zwickau - University of Applied Sciences Zwickau.

**Appendix A.** Link parameters of the Sioux Falls network (All costs in the unit of $)

| i | j | $t_{ij}$ | $c_{ij}$ | $w_{ij}$ | i | j | $t_{ij}$ | $c_{ij}$ | $w_{ij}$ | i | j | $t_{ij}$ | $c_{ij}$ | $w_{ij}$ |
|---|---|---|---|---|---|---|---|---|---|---|---|---|---|---|
| 1 | 2 | 6 | 0 | 25900 | 18 | 16 | 3 | 0 | 19680 | 10 | 9 | 3 | 400 | 13916 |
| 2 | 1 | 6 | 0 | 25900 | 18 | 20 | 4 | 0 | 23403 | 10 | 15 | 6 | 400 | 13512 |
| 3 | 4 | 4 | 0 | 17111 | 19 | 15 | 3 | 0 | 14565 | 11 | 4 | 6 | 400 | 4909 |
| 4 | 3 | 4 | 0 | 17111 | 20 | 18 | 4 | 0 | 23403 | 11 | 14 | 4 | 400 | 4877 |
| 4 | 5 | 2 | 0 | 17783 | 20 | 21 | 6 | 0 | 5060 | 12 | 3 | 4 | 400 | 23403 |
| 5 | 4 | 2 | 0 | 17783 | 20 | 22 | 5 | 0 | 5076 | 12 | 13 | 3 | 400 | 25900 |
| 5 | 6 | 4 | 0 | 4948 | 21 | 20 | 6 | 0 | 5060 | 13 | 12 | 3 | 400 | 25900 |
| 6 | 5 | 4 | 0 | 4948 | 21 | 24 | 3 | 0 | 4885 | 14 | 11 | 4 | 400 | 4877 |
| 7 | 8 | 3 | 0 | 7842 | 22 | 20 | 5 | 0 | 5076 | 14 | 23 | 4 | 400 | 4925 |
| 8 | 7 | 3 | 0 | 7842 | 22 | 23 | 4 | 0 | 5000 | 15 | 10 | 6 | 400 | 13512 |
| 8 | 9 | 10 | 0 | 5050 | 23 | 22 | 4 | 0 | 5000 | 15 | 22 | 3 | 400 | 9599 |
| 9 | 8 | 10 | 0 | 5050 | 24 | 13 | 4 | 0 | 5091 | 16 | 8 | 5 | 400 | 5046 |
| 10 | 11 | 5 | 0 | 10000 | 24 | 21 | 3 | 0 | 4885 | 16 | 17 | 2 | 400 | 5230 |
| 10 | 16 | 4 | 0 | 4855 | 1 | 3 | 4 | 400 | 23403 | 17 | 16 | 2 | 400 | 5230 |
| 10 | 17 | 8 | 0 | 4994 | 2 | 6 | 5 | 400 | 4958 | 17 | 19 | 2 | 400 | 4824 |
| 11 | 10 | 5 | 0 | 10000 | 3 | 1 | 4 | 400 | 23403 | 19 | 17 | 2 | 400 | 4824 |
| 11 | 12 | 6 | 0 | 4909 | 3 | 12 | 4 | 400 | 23403 | 19 | 20 | 4 | 400 | 5003 |
| 12 | 11 | 6 | 0 | 4909 | 4 | 11 | 6 | 400 | 4909 | 20 | 19 | 4 | 400 | 5003 |
| 13 | 24 | 4 | 0 | 5091 | 5 | 9 | 5 | 400 | 10000 | 21 | 22 | 2 | 400 | 5230 |
| 14 | 15 | 5 | 0 | 5128 | 6 | 2 | 5 | 400 | 4958 | 22 | 15 | 3 | 400 | 9599 |
| 15 | 14 | 5 | 0 | 5128 | 6 | 8 | 2 | 400 | 4899 | 22 | 21 | 2 | 400 | 5230 |
| 15 | 19 | 3 | 0 | 14565 | 8 | 6 | 2 | 400 | 4899 | 23 | 14 | 4 | 400 | 4925 |
| 16 | 10 | 4 | 0 | 4855 | 8 | 16 | 5 | 400 | 5046 | 23 | 24 | 2 | 400 | 5079 |
| 16 | 18 | 3 | 0 | 19680 | 9 | 5 | 5 | 400 | 10000 | 24 | 23 | 2 | 400 | 5079 |
| 17 | 10 | 8 | 0 | 4994 | 9 | 10 | 3 | 400 | 13916 | | | | | |

**Appendix B.** OD demand for the Sioux Falls network

| OD ID | Origin | Destination | Demand | OD ID | Origin | Destination | Demand |
|---|---|---|---|---|---|---|---|
| 1 | 2 | 1 | 100 | 16 | 1 | 18 | 100 |
| 2 | 12 | 1 | 200 | 17 | 2 | 18 | 100 |
| 3 | 18 | 1 | 100 | 18 | 12 | 18 | 200 |
| 4 | 13 | 1 | 500 | 19 | 13 | 18 | 100 |
| 5 | 20 | 1 | 300 | 20 | 20 | 18 | 400 |
| 6 | 1 | 2 | 100 | 21 | 1 | 13 | 500 |
| 7 | 12 | 2 | 100 | 22 | 2 | 13 | 300 |
| 8 | 18 | 2 | 100 | 23 | 12 | 13 | 1300 |
| 9 | 13 | 2 | 300 | 24 | 18 | 13 | 100 |
| 10 | 20 | 2 | 100 | 25 | 20 | 13 | 600 |

| | | | | | | | | |
|---|---|---|---|---|---|---|---|---|
| | 11 | 1 | 12 | 200 | 26 | 1 | 20 | 300 |
| | 12 | 2 | 12 | 100 | 27 | 2 | 20 | 100 |
| | 13 | 18 | 12 | 200 | 28 | 12 | 20 | 400 |
| | 14 | 13 | 12 | 1300 | 29 | 18 | 20 | 400 |
| | 15 | 20 | 12 | 500 | 30 | 13 | 20 | 600 |

**Appendix C.** Fixed route link flows, capacities, and fares (links with positive flows only) (Case 1: $\alpha_t = 1$, $\alpha_c = 0.5$).

| Link $l$ | Operator | Flow | Fare ($) | $y_l^*$ | Expected capacity $w_l y_l^*$ | Link $l$ | Operator | Flow | Fare ($) | $y_l^*$ | Expected capacity $w_l y_l^*$ |
|---|---|---|---|---|---|---|---|---|---|---|---|
| (1, 3) | 1 | 376.88 | 0.59 | 0.08 | 376.88 | (16, 8) | 2 | 7.06 | 1.24 | 0 | 7.06 |
| (2, 6) | 2 | 94.54 | 0.56 | 0.02 | 94.54 | (16, 17) | 2 | 0.89 | 0.54 | 0 | 0.89 |
| (3, 1) | 1 | 378.59 | 0.6 | 0.08 | 378.59 | (17, 16) | 2 | 0.87 | 0.57 | 0 | 0.87 |
| (3, 12) | 1 | 441.84 | 0.21 | 0.09 | 441.84 | (17, 19) | 2 | 0.58 | 0.44 | 0 | 0.58 |
| (6, 2) | 2 | 93.53 | 0.34 | 0.02 | 93.53 | (19, 17) | 2 | 0.57 | 1.06 | 0 | 0.57 |
| (6, 8) | 2 | 140.96 | 0.11 | 0.03 | 140.96 | (19, 20) | 2 | 1.74 | 0.9 | 0 | 1.74 |
| (8, 6) | 2 | 138.71 | 0.11 | 0.03 | 138.71 | (20, 19) | 2 | 1.76 | 0.29 | 0 | 1.76 |
| (8, 16) | 2 | 6.84 | 1.04 | 0 | 6.84 | (21, 22) | 4 | 124.29 | 0.71 | 0.02 | 124.29 |
| (12, 3) | 1 | 435.58 | 0.21 | 0.09 | 435.58 | (22, 15) | 4 | 0.68 | 0 | 0 | 0.68 |
| (12, 13) | 1 | 2497.91 | 1.46 | 0.5 | 2497.91 | (22, 21) | 4 | 134.51 | 0.69 | 0.03 | 134.51 |
| (13, 12) | 1 | 2597.84 | 1.44 | 0.52 | 2597.84 | (23, 24) | 3 | 51.46 | 0.96 | 0.01 | 51.46 |
| (15, 22) | 4 | 0.62 | 0.62 | 0 | 0.62 | (24, 23) | 3 | 46.66 | 1.04 | 0.01 | 46.66 |

**Appendix D**. MoD link flows and fares (links with positive flows only) (Case 1: $\alpha_t = 1$, $\alpha_c = 0.5$).

| Link | Operator | Flow | Fare ($) | Link | Operator | Flow | Fare ($) | Link | Operator | Flow | Fare ($) |
|---|---|---|---|---|---|---|---|---|---|---|---|
| (1, 2) | 5 | 60.32 | 1.31 | (12, 10) | 5 | 30.16 | 2.39 | (22, 20) | 6 | 64.21 | 0.16 |
| (1, 3) | 5 | 120.84 | 0.92 | (12, 11) | 5 | 9.61 | 2.18 | (22, 21) | 6 | 10.4 | 1.38 |
| (1, 4) | 5 | 0.03 | 1.87 | (12, 16) | 5 | 92.56 | 2.15 | (22, 23) | 6 | 2.93 | 0.87 |
| (1, 5) | 5 | 0 | 1.53 | (16, 1) | 5 | 9 | 2.15 | (22, 24) | 6 | 17.76 | 1.51 |
| (1, 6) | 5 | 23.21 | 1.13 | (16, 2) | 5 | 11.48 | 1.98 | (23, 13) | 6 | 16.11 | 1.34 |
| (1, 8) | 5 | 39.2 | 1.24 | (16, 6) | 5 | 2.76 | 1.75 | (23, 20) | 6 | 21.72 | 0.44 |
| (1, 11) | 5 | 0 | 2.19 | (16, 8) | 5 | 1.88 | 1.53 | (23, 21) | 6 | 0.48 | 1.24 |
| (1, 12) | 5 | 332.21 | 1.13 | (16, 10) | 5 | 4.01 | 2.21 | (23, 22) | 6 | 2.98 | 0.81 |
| (1, 16) | 5 | 9.2 | 2.15 | (16, 11) | 5 | 20.77 | 2.12 | (23, 24) | 6 | 1.93 | 2.08 |
| (2, 1) | 5 | 59.38 | 1.35 | (16, 12) | 5 | 92.84 | 2.13 | (24, 13) | 6 | 111.6 | 0.9 |

| | | | | | | | | | | |
|---|---|---|---|---|---|---|---|---|---|---|
| (2, 3) | 5 | 71.81 | 0.9 | (13, 15) | 6 | 0.74 | 1.23 | (24, 15) | 6 | 0.07 | 1.84 |
| (2, 4) | 5 | 0.14 | 1.92 | (13, 19) | 6 | 0.69 | 1.66 | (24, 19) | 6 | 0.3 | 1.3 |
| (2, 5) | 5 | 0.17 | 1.54 | (13, 20) | 6 | 340.94 | 1.49 | (24, 20) | 6 | 125.42 | 1.47 |
| (2, 6) | 5 | 44.98 | 0.83 | (13, 21) | 6 | 120.9 | 1.25 | (24, 21) | 6 | 42.43 | 1.32 |
| (2, 8) | 5 | 75.51 | 1.02 | (13, 22) | 6 | 47.25 | 1.49 | (24, 22) | 6 | 15.68 | 1.64 |
| (2, 12) | 5 | 193.55 | 1.16 | (13, 23) | 6 | 14.6 | 1.43 | (24, 23) | 6 | 2.08 | 1.99 |
| (2, 16) | 5 | 10.89 | 2.06 | (13, 24) | 6 | 101.62 | 0.95 | (7, 16) | 7 | 0.11 | 1.29 |
| (3, 1) | 5 | 115.29 | 1.01 | (15, 13) | 6 | 0.69 | 1.91 | (7, 18) | 7 | 61.97 | 1.05 |
| (3, 2) | 5 | 70.84 | 0.9 | (15, 17) | 6 | 0 | 1.12 | (7, 20) | 7 | 24.49 | 1.21 |
| (3, 6) | 5 | 0.06 | 1.75 | (15, 19) | 6 | 0 | 0.85 | (8, 16) | 7 | 2.55 | 1.22 |
| (3, 12) | 5 | 128.04 | 0.54 | (15, 20) | 6 | 0.21 | 0.98 | (8, 17) | 7 | 0.05 | 0.77 |
| (4, 1) | 5 | 0.03 | 1.89 | (15, 22) | 6 | 0 | 1.07 | (8, 18) | 7 | 141.79 | 1.05 |
| (4, 2) | 5 | 0.14 | 1.9 | (15, 24) | 6 | 0.09 | 1.84 | (8, 19) | 7 | 0 | 0.47 |
| (4, 12) | 5 | 0.18 | 1.24 | (17, 15) | 6 | 0 | 1.12 | (8, 20) | 7 | 37.58 | 1.85 |
| (5, 1) | 5 | 0 | 1.95 | (17, 19) | 6 | 0.15 | 0.42 | (16, 7) | 7 | 0.12 | 1.43 |
| (5, 2) | 5 | 0.18 | 1.45 | (17, 20) | 6 | 0.25 | 1.41 | (16, 8) | 7 | 2.59 | 1.45 |
| (5, 12) | 5 | 0.12 | 1.61 | (17, 22) | 6 | 0 | 1.3 | (16, 17) | 7 | 0.12 | 1.01 |
| (6, 1) | 5 | 23.97 | 0.75 | (19, 13) | 6 | 0.67 | 2.26 | (16, 18) | 7 | 61.65 | 0.21 |
| (6, 2) | 5 | 42.9 | 0.69 | (19, 15) | 6 | 0 | 0.83 | (16, 19) | 7 | 0.1 | 1.19 |
| (6, 3) | 5 | 0.07 | 1.45 | (19, 17) | 6 | 0.15 | 1.03 | (16, 20) | 7 | 3.74 | 0.52 |
| (6, 8) | 5 | 18.49 | 0.43 | (19, 20) | 6 | 0.52 | 0.84 | (17, 8) | 7 | 0.05 | 0.99 |
| (6, 12) | 5 | 0.19 | 1.67 | (19, 21) | 6 | 0.09 | 1.76 | (17, 16) | 7 | 0.12 | 0.99 |
| (6, 16) | 5 | 2.76 | 1.5 | (19, 22) | 6 | 0 | 0.93 | (17, 18) | 7 | 0.31 | 0.91 |
| (8, 1) | 5 | 39.9 | 0.88 | (19, 24) | 6 | 0.31 | 1.79 | (17, 19) | 7 | 0.07 | 1.06 |
| (8, 2) | 5 | 76.81 | 0.74 | (20, 13) | 6 | 377.7 | 1.4 | (17, 20) | 7 | 0.14 | 1.54 |
| (8, 6) | 5 | 18.38 | 0.43 | (20, 15) | 6 | 0.15 | 0.98 | (18, 7) | 7 | 62.08 | 1.33 |
| (8, 12) | 5 | 2.96 | 0.95 | (20, 17) | 6 | 0.27 | 1.29 | (18, 8) | 7 | 140.74 | 1.33 |
| (8, 16) | 5 | 1.87 | 1.33 | (20, 19) | 6 | 0.5 | 0.24 | (18, 16) | 7 | 61.71 | 0.21 |
| (10, 11) | 5 | 4.59 | 2.44 | (20, 21) | 6 | 99.31 | 1.2 | (18, 17) | 7 | 0.3 | 0.93 |
| (10, 12) | 5 | 29.89 | 2.39 | (20, 22) | 6 | 69.77 | 0.16 | (18, 19) | 7 | 0.2 | 1.27 |
| (10, 16) | 5 | 4.06 | 2.2 | (20, 23) | 6 | 24.68 | 0.44 | (18, 20) | 7 | 92.64 | 1.85 |
| (11, 1) | 5 | 0 | 2.19 | (20, 24) | 6 | 137 | 1.41 | (19, 8) | 7 | 0 | 1.09 |
| (11, 10) | 5 | 4.63 | 2.44 | (21, 13) | 6 | 131.05 | 1.21 | (19, 16) | 7 | 0.1 | 1.79 |
| (11, 12) | 5 | 9.52 | 2.18 | (21, 19) | 6 | 0.07 | 1.74 | (19, 17) | 7 | 0.06 | 1.7 |
| (11, 16) | 5 | 20.34 | 2.18 | (21, 20) | 6 | 90.91 | 1.25 | (19, 18) | 7 | 0.21 | 1.8 |
| (12, 1) | 5 | 333.86 | 1.14 | (21, 22) | 6 | 8.96 | 1.58 | (19, 20) | 7 | 0.82 | 0.84 |
| (12, 2) | 5 | 196.78 | 1.11 | (21, 23) | 6 | 0.49 | 1.24 | (20, 7) | 7 | 23.91 | 1.51 |
| (12, 3) | 5 | 129.52 | 0.52 | (21, 24) | 6 | 46.01 | 1.28 | (20, 8) | 7 | 38.3 | 2.07 |
| (12, 4) | 5 | 0.19 | 1.32 | (22, 13) | 6 | 50.09 | 1.48 | (20, 16) | 7 | 3.86 | 0.52 |
| (12, 5) | 5 | 0.12 | 1.75 | (22, 15) | 6 | 0 | 1.09 | (20, 17) | 7 | 0.14 | 1.58 |
| (12, 6) | 5 | 0.18 | 1.95 | (22, 17) | 6 | 0 | 1.44 | (20, 18) | 7 | 93.24 | 1.84 |
| (12, 8) | 5 | 3.14 | 1.13 | (22, 19) | 6 | 0 | 0.93 | (20, 19) | 7 | 0.8 | 0.24 |

**Appendix E**. Walking link flows (links with positive flows only) (Case 1: $\alpha_t = 1$, $\alpha_c = 0.5$).

| Link | Flow | Link | Flow | Link | Flow |
|---|---|---|---|---|---|
| (1, 2) | 193.21 | (10, 16) | 73.58 | (20, 18) | 441.84 |
| (2, 1) | 195.14 | (11, 10) | 42.85 | (20, 21) | 310.45 |
| (3, 4) | 0.43 | (11, 12) | 58.3 | (20, 22) | 180.01 |
| (4, 3) | 0.39 | (12, 11) | 58.22 | (21, 20) | 290.04 |
| (4, 5) | 0.39 | (13, 24) | 474.45 | (21, 24) | 372.69 |
| (5, 4) | 0.35 | (15, 19) | 0.93 | (22, 20) | 165.75 |
| (5, 6) | 0.34 | (16, 10) | 72.92 | (22, 23) | 39.11 |
| (6, 5) | 0.3 | (16, 18) | 161.94 | (23, 22) | 35.87 |
| (7, 8) | 86.11 | (18, 16) | 162.48 | (24, 13) | 513.14 |
| (8, 7) | 86.58 | (18, 20) | 442.91 | (24, 21) | 345.94 |
| (10, 11) | 42.45 | (19, 15) | 0.9 | | |

**Appendix F**. MoD zone capacities (operated zones only) (Case 1: $\alpha_t = 1$, $\alpha_c = 0.5$).

| Zone $i$ | Operator | $v_i^*$ | Expected Capacity | Zone $i$ | Operator | $v_i^*$ | Expected Capacity |
|---|---|---|---|---|---|---|---|
| 1 | 5 | 0.12 | 585.02 | 17 | 6 | 0 | 0.41 |
| 2 | 5 | 0.09 | 456.43 | 19 | 6 | 0 | 1.74 |
| 3 | 5 | 0.06 | 314.23 | 20 | 6 | 0.14 | 709.39 |
| 4 | 5 | 0 | 0.35 | 21 | 6 | 0.06 | 277.5 |
| 5 | 5 | 0 | 0.3 | 22 | 6 | 0.03 | 145.39 |
| 6 | 5 | 0.02 | 88.38 | 23 | 6 | 0.01 | 43.22 |
| 8 | 5 | 0.03 | 139.93 | 24 | 6 | 0.06 | 297.59 |
| 9 | 5 | 0 | 0 | 7 | 7 | 0.02 | 86.58 |
| 10 | 5 | 0.01 | 38.55 | 8 | 7 | 0.04 | 181.98 |
| 11 | 5 | 0.01 | 34.5 | 16 | 7 | 0.01 | 68.32 |
| 12 | 5 | 0.16 | 796.11 | 17 | 7 | 0 | 0.69 |
| 16 | 5 | 0.03 | 142.76 | 18 | 7 | 0.07 | 357.68 |
| 13 | 6 | 0.13 | 626.74 | 19 | 7 | 0 | 1.19 |
| 14 | 6 | 0 | 0 | 20 | 7 | 0.03 | 160.25 |
| 15 | 6 | 0 | 1 | | | | |